\documentclass[a4paper,10pt,aps,prd,showpacs,preprintnumbers,twocolumn,nofootinbib,superscriptaddress]{revtex4-2}

\usepackage[english]{babel}
\usepackage[utf8]{inputenc}
\usepackage[T1]{fontenc}
\usepackage{newtxtext,newtxmath}
\usepackage{color}
\usepackage{url}
\usepackage{hyperref}
\usepackage{xcolor}
\usepackage{graphicx}
\usepackage{float}
\usepackage{amsmath}
\usepackage{mathrsfs}
\usepackage{verbatim}
\usepackage{booktabs}
\usepackage{multirow}
\usepackage{lettrine}
\usepackage{natbib}

\def \be {\begin{equation}}
\def \ee {\end{equation}}
\def \dd {\mathrm{d}} 
\def \t {\tilde}
\def \p {\partial}
\def \l {\left}
\def \r {\right}
\def \bs {\boldsymbol}
\def \dd {\mathrm{d}} 
\def \bs {\boldsymbol}

\newcommand{\e}[1]{_{\rm #1}}

\definecolor{citeco}{HTML}{2E3093}
\definecolor{linksco}{HTML}{2E3093}

\begin{document}

\title{Redshift tomography of the kinematic matter dipole}
\author{Sebastian von Hausegger}
\email{sebastian.vonhausegger@physics.ox.ac.uk}
\affiliation{Department of Physics, University of Oxford, Parks Road, Oxford OX1 3PU, United Kingdom}
\author{Charles Dalang}
\email{c.dalang@qmul.ac.uk}
\affiliation{School of Mathematical Sciences, Queen Mary University of London, \\
Mile End Road, London E1 4NS, United Kingdom}
\affiliation{Institute of Cosmology and Gravitation, University of Portsmouth, \\
Burnaby Road, Portsmouth PO1 3FX, United Kingdom}

\begin{abstract}
The dipole anisotropy induced by our peculiar motion in the sky distribution of cosmologically distant sources is an important consistency test of the standard FLRW cosmology. In this work, we formalize how to compute the kinematic matter dipole in redshift bins.  Apart from the usual terms arising from angular aberration and flux boosting, there is a contribution from the boosting of the redshifts that becomes important when considering a sample selected on observed redshift, leading to non-vanishing correction terms.  We discuss examples and provide expressions to incorporate arbitrary redshift selection functions.  We also discuss the effect of redshift measurement uncertainties in this context, in particular in upcoming surveys for which we provide estimates of the correction terms. Depending on the shape of a sample's redshift distribution and on the applied redshift cuts, the correction terms can become substantial, even to the degree that the direction of the dipole is reversed.  Lastly, we discuss how cuts on variables correlated with observed redshift, such as color, can induce additional correction terms.
\end{abstract}

\maketitle

\section{Introduction}
\label{sec:introduction}

\lettrine{T}{he} foundational assumptions of the standard cosmological model are homogeneity and isotropy, also known as the Cosmological Principle, which together motivate the use of the Friedmann-Lemaître-Robertson-Walker (FLRW) spacetime.  In such a spacetime, all comoving observers would see isotropic radiation and matter fields~\citep{2010fimv.book..267S}, which in turn would be sufficient to prove that we live in a Friedmann universe~\citep{Ehlers:1966ad,Stoeger:1994qs}.  However, firstly, we only perform observations on or within our past light cone, so only from one vantage point; secondly, we are \textit{not} comoving observers; thirdly, homogeneity and isotropy are only expected to be satisfied statistically, in practice. Tests of the Cosmological Principle therefore become consistency checks\footnote{For recent reviews see, e.g.,~\cite{Peebles:2022akh,Aluri:2022hzs}.}, and, having entered the era of big data in cosmology, it is now possible to undertake such tests with observations rather than simply going along with the above assumptions.  An important question in particular is whether the reference frames in which radiation and matter appear statistically isotropic are truly the same, as is the case in the standard model.

The rest frame of radiation can be defined as the frame in which the Cosmic Microwave Background (CMB) appears statistically isotropic.  This is expected to coincide with the frame in which the CMB dipole, considered to be the Doppler-boosted monopole~\cite{1967Natur.216..748S}, vanishes.  To-date, the highest-precision measurement of the CMB dipole was performed by the Planck Collaboration, who reported an amplitude of $\Delta T=(3.36208\pm0.00099)\,{\rm mK}$, corresponding to a boost\footnote{$\beta$ denotes the observer speed $v$ relative to $c$, the speed of light, $\beta\equiv v/c$.} of $\beta=(1.2336\pm0.0037)\times10^{-3}$ towards $(l,b)=(264.021\pm0.011^\circ, 48.253\pm0.005^\circ)$~\cite{Planck:2018nkj}. A number of tests for consistency of this measurement with its kinematic interpretation are possible using CMB data alone.  For example, measurements of the dipolar modulation of the higher multipole temperature (and polarization) fluctuations, expected from special relativistic aberration~\cite{Challinor:2002zh}, generally agree in amplitude and direction with that of the dipole itself~\cite{Planck:2013kqc,Ferreira:2020aqa,Saha:2021bay}.  While such measurements, within measurement uncertainties, support the kinematic origin hypothesis, they do not exclude other scenarios, such as those in which large-scale isocurvature perturbations contribute to the CMB dipole~\cite{1988ASPC....4..344G,Turner:1991dn,Langlois:1995ca,Erickcek:2008jp,Domenech:2022mvt}.  Therefore it is also necessary to test whether there is consistency with the rest frame of matter.

The rest frame of matter is determined by observing tracers of the matter distribution, such as galaxies at high redshift, where the Universe is expected to be approximately homogeneous.  If such a sample is indeed isotropic across the sky for a comoving observer,~\citet{1984MNRAS.206..377E} noted that an observer moving with respect to the matter rest frame should see a dipolar modulation of source number counts in a flux-limited catalog, directly proportional to their boost $\beta$.  A cut on observed flux $S>S_*$ selects more sources in the direction of motion in which fluxes were Doppler boosted upwards.  This dipole is predicted to have an amplitude $\mathcal{D}_{\rm kin}$ that depends only on observed properties of the given sample: the power law index $x$ of the projected integral source counts $\mathcal{N}(S>S_*)\propto S_*^{-x}$ and the average spectral index $\alpha$ of the source spectra $S_\nu\propto\nu^{-\alpha}$, both evaluated directly at the flux cut $S_*$, and the observer's boost $\beta$,  
\begin{align}
    \mathcal{D}_{\rm kin}=\left[2+x(1+\alpha)\right]\beta\,.\label{eq:DEB}
\end{align}
The same special relativistic effects that create a dipole in the CMB are responsible for this matter dipole: the factor of $2$ arises from angular aberration that scales the unit solid angle of the number density, while the additional terms arise due to the Doppler boosting of the sources' flux above and below the flux cut.  Moreover, similarly to the aberration of CMB fluctuations, an analogous computation predicts a concomitant dipolar modulation of the number count fluctuations~\cite{Lacasa:2024ybp}.  Hence just as with the CMB dipole, the matter rest frame can be defined with observations of the matter dipole.

Our ability to measure the matter dipole has however been limited by the availability of sufficiently large high-redshift galaxy catalogs. While radio catalogs were considered the best-suited for this measurement, due to radio AGN (active galactic nuclei) predominantly occupying redshifts $z\sim1$, the necessary procedures of catalog cleaning suppressed the eligible number of sources to $\mathcal{O}(10^5)$ where shot noise undermines a precise dipole measurement.\footnote{For comparison, a significant measurement of the expected matter dipole is achieved only with $\mathcal{O}(10^6)$ sources.~\cite{Crawford:2008nh,Secrest:2025bcw}}  Nevertheless, early studies employing the $1.4\,{\rm GHz}$ NRAO VLA Sky Survey (NVSS) catalog, resulted in observed dipoles that agreed with the CMB dipole expectation in direction, but not in amplitude~\cite{2002Natur.416..150B,2011ApJ...742L..23S,2013A&A...555A.117R}.  Since the discrepancy was a mild $\sim2-3\sigma$, more precise measurements were required to conclusively test the consistency between radiation and matter rest frames. Only recently, this requirement was addressed via the new CatWISE catalog of high-redshift quasars in the mid-infrared, compiled using data from the WISE survey. This was the first significant measurement of the matter dipole~\cite{Secrest:2020has,Secrest:2022uvx} and it confirmed the discrepancy in amplitude previously reported with radio observations. Subsequent independent confirmation with alternative methods and more data now places the `matter dipole anomaly' on firm grounds with a significance of over $5\sigma$~\cite{Secrest:2022uvx,Dam:2022wwh,Wagenveld:2023kvi,Oayda:2024hnu}. Intriguingly, the rest frames of radiation and matter appear \textit{not} to coincide. The explanation for this anomaly remains unknown.

Tomographic measurements in redshift appear to be a natural next step to the previous redshift-integrated studies.  Such measurements might reveal a redshift-dependence of the matter dipole amplitude that could help in finding the explanation for this tension.  However, selection in observed redshift requires additional considerations to those made for Eq.\,\eqref{eq:DEB}: Redshifts of sources are boosted according to
\begin{align}
    (1+z) = (1+\bar z)(1-\beta\cos\theta)\,,\label{eq:z_boost}
\end{align}
to linear order in $\beta$, where $z$ is the observed (boosted) redshift of a source, $\bar z$ is the cosmological redshift as seen by a comoving observer, and $\theta$ is the angle between the source and the boost direction. A sharp cut on \textit{observed} redshift $z$ would provoke an apparent dipole anisotropy in a previously isotropic catalog in the same manner as above.  The sharp cut on observed redshift would correspond to lower cosmological redshifts in the direction of motion, and conversely in the opposite direction --- even in the absence of the above aberration and Doppler boosting of the sources' measured fluxes, Eq.~\eqref{eq:DEB}. Note that a cut on \textit{cosmological} redshift $\bar{z}$ would \textit{not} lead to such a dipole anisotropy.

Indeed, the possibility of a redshift-dependent kinematic matter dipole was introduced in Ref.~\cite{Maartens:2017qoa}, which suggested that general relativistic effects are important for the correct determination of the expected dipole amplitude, when expressed in terms of the sources' luminosity function instead of the observed sample properties $x$ and $\alpha$ of Eq.~\eqref{eq:DEB}. In subsequent work, connection between this approach and Eq.~\eqref{eq:DEB} was sought~\cite{Nadolny:2021hti,Dalang:2021ruy,Domenech:2022mvt,Guandalin:2022tyl}, before their exact equivalence was demonstrated generally~\cite{vonHausegger:2024jan}.  This equivalence depends however on assuming that the source number distribution with redshift vanishes at the integration boundaries.  Dropping this assumption results in an additional contribution to the matter dipole amplitude.

In this work we quantify how to compute this contribution and prepare for the opportunities that will come with near-future surveys of the large scale structure.  After a brief theoretical overview in Sec.~\ref{sec:theory}, we provide a first example of the added contribution to the kinematic matter dipole that boundary terms can provide.  Sec.~\ref{sec:redshiftselectionfunctions} covers a more general treatment of the influence of boundary terms, including general selection functions and redshift uncertainties.  In Sec.~\ref{sec:practicalexamples}, we discuss strategies to estimate the relevant quantities and make predictions for upcoming surveys. We discuss and conclude in Secs.~\ref{sec:discussion} and~\ref{sec:conclusion}.

\section{Theoretical Foundation}
\label{sec:theory}

First, we review different theoretical expressions for the redshift-dependent dipole, discuss their respective advantages and how they relate to one another.  The source number count dipole as a function of redshift, derived on a FLRW background, reads
\cite{Maartens:2017qoa}
\begin{align}
\mathcal{D}\e{kin}(z) = \l[2 +\frac{\dot{\mathcal{H}}(z)}{\mathcal{H}^2(z)} + \frac{2(1-x(z))}{r(z)\mathcal{H}(z)} - b\e{e}(z) \r]\beta\,.  \label{eq:Dz1}
\end{align}
The factor of $2$ comes from aberration of solid angles, $\mathcal{H}$ denotes the Hubble rate $\mathcal{H} = \dot a a^{-1}$ and dots indicate derivatives with respect to conformal time $\dd \eta = \dd t/a$ and $t$ indicates cosmic time. The comoving distance as a function of redshift is written as $r(z)$.  The magnification and evolution biases, $x$ and $b\e{e}$, are defined as
\begin{align}
    x(z) & \equiv -\frac{\partial \log n(z,S_*)}{\partial \log S_*} 
    \label{eq:x}\,,\\
    b\e{e}(z) & \equiv - \frac{\p \log n\e{c}(z,S_*)}{\p \log(1+z)} \,,
\end{align}
where $n\e{c}(z,S_*)= \dd N/\dd V\e{c}(z,S_*)$ indicates the comoving source number density and $n(z,S_*)\equiv{\rm d}N/{\rm d}\Omega{\rm d}z\,(z,S>S_*)$.  These quantities should be understood as the average (or background) number densities which can be estimated from the monopole of the number counts.  The disadvantage of Eq.~\eqref{eq:Dz1} is that it is cosmology-dependent.  In particular, the evolution bias is challenging to evaluate~\citep{Wang:2020ibf}. It was first shown in App.\,A of~\cite{Nadolny:2021hti} that if the luminosities of the sources obey a power law in frequency\footnote{Note that it does not matter if $\alpha$ evolves with cosmological redshift, so long as it is measured on the sources which are close to the flux limit~\cite{vonHausegger:2024jan}. Likewise, if the proportionality factor $\kappa$ evolves with cosmological redshift, this only modifies the monopole of the source number counts and leaves the dipole unaffected.} $L = \kappa \nu^{-\alpha}$ with spectral index $\alpha$, then the redshift-dependent kinematic matter dipole amplitude can alternatively be written as
\begin{align}
    \mathcal{D}_{\rm kin}(z) = \left[3+x(z)\left(1+\alpha(z)\right)+\frac{{\rm d}\log n(z)}{{\rm d}\log (1+z)}\right]\beta\,,\label{eq:Dz2}
\end{align}
where $z$ is the observed redshift. If integrated over redshift weighted by the distribution of sources $n(z)$, the last term can be integrated by parts, such that, if $n(z)$ vanishes at the redshift boundaries of the sample, the equivalence with the Ellis and Baldwin formula follows~\citep{Nadolny:2021hti} as we also show below.  The requirement that boundary terms vanish is typical if one observes all sources independently of their distance to the observer.  Instead, if one divides the sample in redshift bins, these boundary terms can become substantial.

\subsection{Boundary terms}
\label{sec:theory_boundaryterms}

In a given redshift range $z\in[z_1,z_2]$, denoted by bin $b$, the kinematic matter dipole expected in a projected and flux-limited sample of galaxies is computed as
\begin{align}
    \mathcal{D}_{\rm kin}(z_b) = \int_{z_1}^{z_2}{\rm d}z\,\mathcal{D}_{\rm kin}(z)f_b(z)\,,\label{eq:Dz_int_tomography}
\end{align}
where the integral source counts are normalised within the same bin as
\begin{align}
    f_b(z) \equiv \frac{n(z)}{\int_{z_1}^{z_2}{\rm d}z\,n(z)}\,.\label{eq:fz_tomography}
\end{align}
We emphasize that $f_b(z)$ is a distribution readily given by the data if the redshifts $z$ are known, i.e.~it does not require the computation of the luminosity function or the luminosities of the sources.  The argument $z_b$ denotes the effective redshift of the redshift bin under investigation and is computed as the redshift $z$ averaged over $f_b(z)$,
\begin{align}
    z_b \equiv \int_{z_1}^{z_2}{\rm d}z\,z\,f_b(z)\,.\label{eq:zeff_tomography}
\end{align}
It is instructive to express the dipole amplitude $\mathcal{D}_{\rm kin}(z)$ in Eq.~\eqref{eq:Dz_int_tomography} by Eq.~\eqref{eq:Dz2}.  This allows a simplification, analogous to the derivation in Ref.~\cite{vonHausegger:2024jan}, which reduces the first part of the integral over Eq.~\eqref{eq:Dz2} to
\begin{align}
    &\int_{z_1}^{z_2}{\rm d}z\,\left[3+x(z)(1+\alpha(z))\right]f_b(z) =\,3+\tilde x_b(1+\tilde \alpha_b)\,,\label{eq:int_part1}
\end{align}
resembling the form of Eq.~\eqref{eq:DEB} (see~Appendix~\ref{app:derivation1} for a derivation respecting more general selection functions).  The last part of the same integral gives
\begin{align}
    &\int_{z_1}^{z_2}{\rm d}z\,\frac{{\rm d}\log n(z)}{{\rm d}\log (1+z)}f_b(z) \nonumber\\
    =&  \int_{z_1}^{z_2}{\rm d}z\,(1+z)\frac{{\rm d}n(z)}{{\rm d}z}\bigg/\int_{z_1}^{z_2} {\rm d}z\,n(z) \,,\label{eq:int_part2}
\end{align}
where we used the definition of $f_b(z)\propto n(z)$.
Finally, we integrate by parts to obtain
\begin{align}
    &\left[(1+z)n(z)\big|_{z_1}^{z_2}-\int_{z_1}^{z_2} {\rm d}z\,n(z)\right]\bigg/\!\int_{z_1}^{z_2} {\rm d}z\,n(z)\nonumber\\
    =& \,(1+z)\,f_b(z)\big|_{z_1}^{z_2}-1\,.\label{eq:int_part2_eval}
\end{align}
In summary, Eq.~\eqref{eq:Dz_int_tomography} together with Eq.~\eqref{eq:Dz2} reduces to
\begin{align}
    \mathcal{D}_{\rm kin}(z_b) = \Big[2+\tilde x_b(1+\tilde \alpha_b)+(1+z)\,f_b(z)\big|_{z_1}^{z_2}\Big]\beta\,.  \label{eq:Dkin_tomography}
\end{align}
Thus, the kinematic matter dipole for a sample cut in redshift receives a contribution from the boundaries of the redshift distribution at $z_1$ and $z_2$, which act as thresholds for sources to move into and out of the investigated sample, analogously and in addition to the effects at the flux threshold $S_*$.  

Before we continue to study the magnitude of these boundary terms, it is worth observing relations between the derived expression and the notorious Eq.~\eqref{eq:DEB}.  For instance, if the redshift distribution fulfills $(1+z_1)f_b(z_1)=(1+z_2)f_b(z_2)$ at the boundaries, Eq.~\eqref{eq:Dkin_tomography} reduces entirely to the form of Eq.~\eqref{eq:DEB}.  This is the case in particular if the redshift distribution simply vanishes at the boundaries $f_b(z_1)=f_b(z_2)=0$, which is mostly appropriate to assume when dealing with full galaxy samples as opposed to sub-samples in redshift slices.\footnote{Indeed Refs.~\cite{Nadolny:2021hti,Dalang:2021ruy,Domenech:2022mvt} had rightfully made this assumption in describing the galaxy samples of e.g.~\citet{Secrest:2020has}.}  Moreover, one typically expects dipole amplitudes \textit{\`a la} Eq.~\eqref{eq:DEB} to be positive.  In this case, however, even if the contributions by $\tilde x$ and $\tilde\alpha$ predict a positive dipole amplitude, the boundary terms may render this expectation negative, i.e.~a dipole pointed opposite to the direction of motion.\footnote{For comparison, in Eq.~\eqref{eq:Dz1}, a negative contribution to the dipole amplitude may come from a positive evolution bias and/or from a magnification bias which satisfies $x(z)>1$.}  Lastly, even if the boundary terms do not vanish, we would like to emphasize that they can simply be read-off the data without reference to underlying luminosity functions, rendering Eq.~\eqref{eq:Dkin_tomography} an observationally entirely tractable quantity.

\subsection{Examples}
\label{sec:theory_examples}

\begin{figure}
    \centering
    \includegraphics[width=0.99\columnwidth]{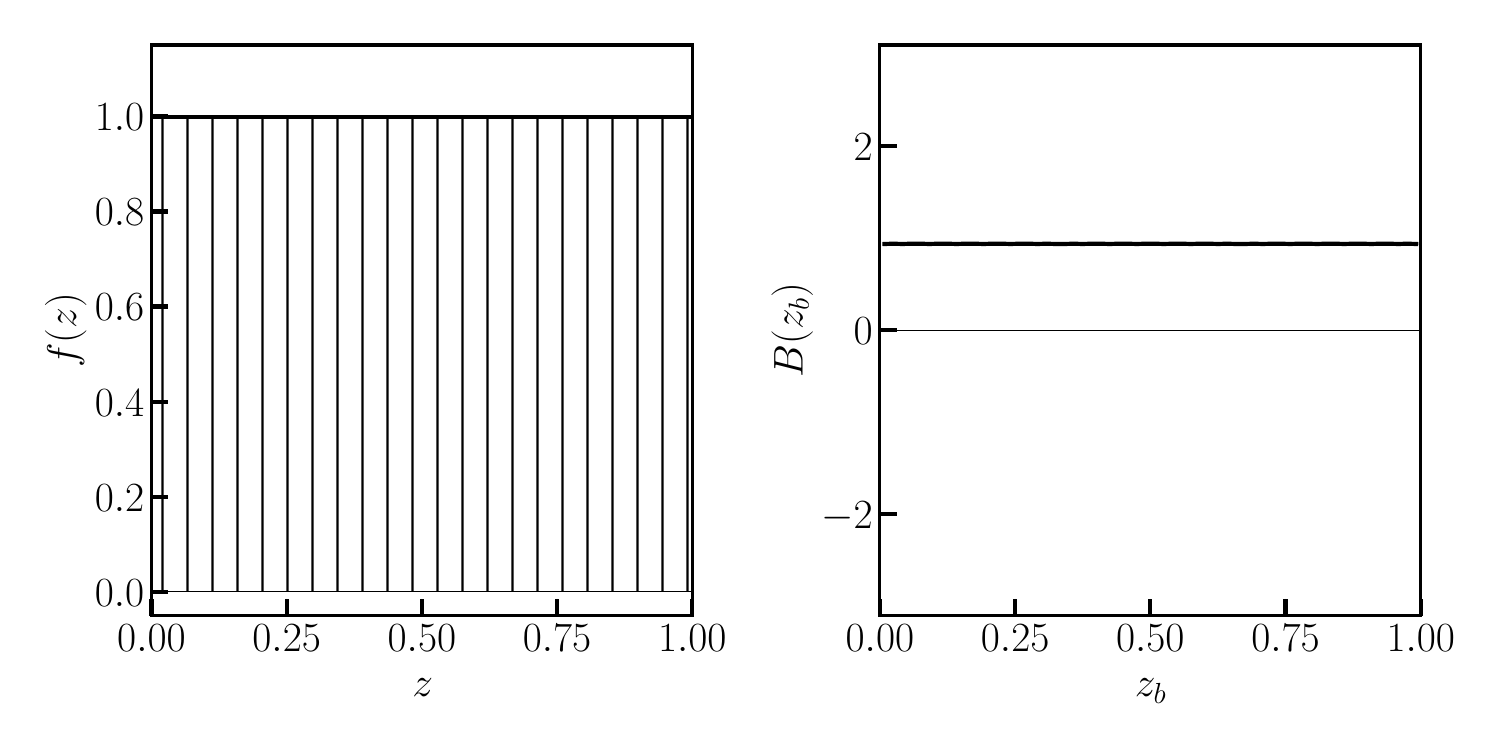}
    \includegraphics[width=0.99\columnwidth]{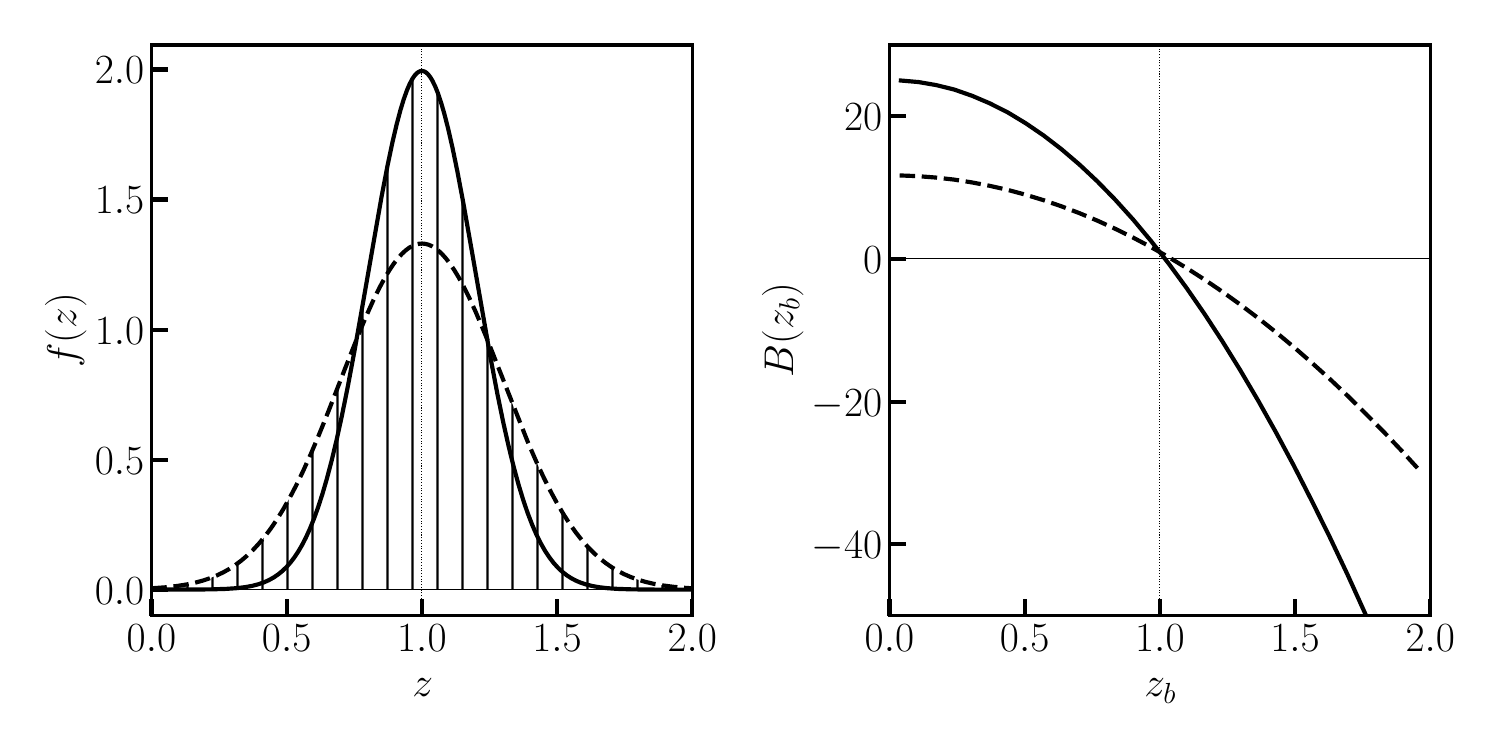}
    \caption{Computation of boundary terms, Eq.~\eqref{eq:B_tomography}, for two different shapes of $f(z)$.  \textit{Top left:} $f(z)$ is chosen to be constant.  \textit{Bottom left:} $f(z)$ is chosen to be Gaussian, solid and dashed lines differ only in their standard deviation.  \textit{Right panels:} $B(z_b)$ corresponding to the distributions $f(z)$ on their left.  Vertical hatching indicates equidistant equal-width top hat selections within the respective distributions.}
    \label{fig:boundary1}
\end{figure}

To get a feeling for the magnitude of the boundary terms in Eq.~\eqref{eq:Dkin_tomography}, we isolate them as
\begin{align}
    B(z_b)\equiv (1+z)\,f_b(z)\big|_{z_1}^{z_2}\,,\label{eq:B_tomography}
\end{align} 
and compute them for two simple examples of $f(z)$, see Fig.~\ref{fig:boundary1}.  The first example, $f(z)=\hbox{const.}$ is trivial in that the normalized $f_b(z)=(z_2-z_1)^{-1}$ simply returns $B=1$.  The second example, for which we choose $f(z)$ to be one of two Gaussians, is presented in the bottom panels.  It can be shown that, for sufficiently thin, equidistant, evenly-spaced top hat selection functions as those considered here, and a given $f(z)$, $B$ is well-approximated by a function of only the effective redshift $z_b$ of the selected bin $b$, not just on its explicit boundaries $z_1$ and $z_2$.  While it simplifies the presentation here, this is not generally the case.  We return to this point below in the context of survey-specific redshift distributions and offer an explanation in App.~\ref{app:moreparameters_bin}.

Whilst it is important to note that $B(z_b)$ are still to-be-multiplied by $\beta$ before contributing to the full dipole amplitude $\mathcal{D}_{\rm kin}(z_b)$, in Eq.~\eqref{eq:Dkin_tomography}, it can be seen that the boundary terms might, in principle, have a substantial impact compared to the Ellis \& Baldwin expectation that for present data typically lies around $\sim 4\beta$. For instance, a constant redshift distribution $f(z)$, leads to an equally constant contribution of $\beta$ to the kinematic matter dipole across effective redshifts $z_b$. For typical values of $\tilde x$ and $\tilde\alpha$ of order unity each, this corresponds to an $\mathcal{O}(20\%)$ contribution to $\mathcal{D}_{\rm kin}$. A general redshift distribution, as exemplified here by Gaussians of different width, can produce boundary terms that would significantly dominate the contributions of $\tilde x$ and $\tilde\alpha$; in other words, the Doppler boosting of redshifts past the boundaries of $f$ largely outweighs the contributions of aberration and Doppler boost of source flux.\\

These introductory examples cover the case where redshifts are measured without uncertainty, approximated most closely by a spectroscopic redshift survey.  In such a case, top hat selection functions can be applied to data and compared directly to the theoretical expressions here. Substantial redshift uncertainties, such as in photometric surveys however, require more detailed considerations that we discuss in the following section, where we present results for general selection functions.

\section{Redshift selection functions}
\label{sec:redshiftselectionfunctions}

In the  previous section, we have calculated the boundary terms for redshift bins, which select observed redshifts with a sharp upper and lower cut-off, assuming vanishing redshift measurement uncertainties.  In practice, redshift tomography of the kinematic dipole will be attempted also with photometric surveys.  While this requires additional considerations, photometric surveys contain up to $\mathcal{O}(10^2)$ times more sources than spectroscopic surveys, reducing the error uncertainties on the matter dipole by a factor of 10.  Sharp cut-offs on photometric redshifts correspond to skewed Gaussian distributions of true redshifts falling to zero in a smooth manner, suggesting that the boundary terms vanish. In fact, we will show that this is incorrect. As suggested by the redshift logarithmic derivative of $n(z)$ in Eq.~\eqref{eq:Dz2} that feeds into Eq.~\eqref{eq:Dz_int_tomography}, it is important to capture correctly the shape of the true redshift distribution of sources and the selection function applied to those. These quantities may not be directly observable, but can be estimated, given relevant data, as we discuss in the following.

It is convenient to cover general selection functions in the context of measurement uncertainties on redshifts.  Photometric redshifts, for instance, are inferred via a number of methods from broad-band measurements, where the recording instrument's spectral resolution is not high enough to detect spectral features, or where the targeted sources do not offer such features for spectroscopic redshifts to be drawn.  Spectroscopic measurements, on the other hand, provide near-exact estimates of the sources' redshifts, and so cross-correlating subsamples of photometric and spectroscopic catalogs can serve as a means to estimate the magnitude of photometric redshift uncertainties.  This procedure can become quite complex.\footnote{Indeed, it forms a field of research on its own (see, for example~\cite{Benitez:2000,Ilbert:2006,Battisti:2019}).}  However, it can be reduced to learning a conditional distribution between photometric and spectroscopic redshifts which subsequently is used for the computation of the true (spectroscopic) redshift distribution given the distribution of uncertain (photometric) redshifts.  In the following we assume that the learning process has already been completed, and so we focus only this last step.

Consider redshift uncertainties such that the boosted redshifts $z$ are actually observed as $z'$.  The respective redshift distributions are related via marginalizing over a conditional distribution as
\begin{align}
    n'(z') = \int_0^\infty{\rm d}z\,n(z)P(z',z)\,,\label{eq:nprime_convolution_main}
\end{align}
where the conditional distribution $P(z',z)$ can be learned with appropriate data as indicated above, and where $n'(z')$ describes the observed distribution of uncertain redshifts $z'$.  Through $P(z',z)$ a given selection of redshifts $z'$ corresponds to a selection of redshifts $z$ (see appendix \ref{app:derivation2} for details).  Denoting this $z'$-selection by bin $b$, we define the selection function of redshifts $z$ as
\begin{align}
    W_b(z) \equiv \int_0^\infty{\rm d}z'\,W'_b(z')P(z',z)\,,\label{eq:Wb_main}
\end{align}
with which we define the distribution of true redshifts $z$ of those sources contained in the $z'$-bin $b$:
\begin{align}
    n_b(z) \equiv n(z)W_b(z)\,.\label{eq:nb_main}
\end{align}
It should be emphasized that, even though we defined $W_b(z)$ as per the relation between redshifts $z'$ and $z$, it can serve also as a general selection function of the sources in $n(z)$, beyond the simple top hat selection discussed in Sec.~\ref{sec:theory_boundaryterms}.  Below, we will consider a specific case in which $W'_b(z')$ are top hat selection functions on redshift $z'$ and $W_b(z)$ are the corresponding selection functions on redshift $z$ as defined by Eq.~\eqref{eq:Wb_main}.

To summarize, for a photometric survey, where the conditional probability $P(z',z)$ has been learned from cross matching with a spectroscopic survey, the observable quantities are the boosted photometric redshift distribution $n'(z')$ and an estimate of the boosted spectroscopic redshift distribution $n(z)$, via the spectroscopic subsamble.  A choice on the photometric selection function  $W'_b(z')$ allows the determination of $W_b(z)$ via Eq.~\eqref{eq:Wb_main}.  These can be used to estimate the binned number count via Eq.~\eqref{eq:nb_main}, which together with $W_b(z)$ represent the core ingredients for estimating the generalization of the boundary terms, as we will see in the next section.

\subsection{Selection terms}
\label{sec:redshiftselectionfunctions_}

Having defined the selection function $W_b(z)$, we are ready to define quantities equivalent to those in Sec.~\ref{sec:theory_boundaryterms}
\begin{align}
    f_b(z) \equiv \frac{n_b(z)}{\int_0^\infty{\rm d}z\,n_b(z)}\,,
    &&
    z_b \equiv \int_0^\infty{\rm d}z\,z\,f_b(z)\,,
    \label{eq:fz_zb_W}
\end{align}
where $n_b(z)$ is the distribution of redshifts $z$ of those objects selected by $W_b(z)$, or equivalently, by $W'_b(z')$ on $n'(z')$, Eq.~\eqref{eq:nb_main}.  As before, the dipole in bin $b$ is computed via
\begin{align}
    \mathcal{D}_{\rm kin}(z_b) = \int_0^\infty{\rm d}z\,\mathcal{D}_{\rm kin}(z)f_b(z)\,,\label{eq:Dz_int_W}
\end{align}
where we input as the kinematic redshift matter dipole the expression of Eq.~\eqref{eq:Dz2}.  A calculation equivalent to that of Sec.~\ref{sec:theory_boundaryterms} brings with it a consideration that we will generalise below in section~\ref{sec:redshiftselectionfunctions_s*dependence}:  for now, we assume the selection function to be independent of flux cut $\partial W_b/\partial S_*=0$.  Having assumed this, we find
\begin{align}
    \mathcal{D}_{\rm kin}(z_b) & = \bigg[2+\tilde{x}_b (1+\tilde\alpha_b) + f_b(z)(1+z)\big|_0^\infty\nonumber \\
    &\quad - \int_0^\infty{\rm d}z\,f_b(z) \frac{{\rm d}\log W_b(z)}{{\rm d}\log (1+z)}\bigg]\,\beta\,,\label{eq:Dkin_W}
\end{align}
where $\tilde x_b$ and $\tilde\alpha_b$ are the values $x$ and $\alpha$ from Eq.\,\eqref{eq:DEB} computed from only the sources selected by bin $b$; see App.\,\ref{app:derivation1} for a full definition and derivation of Eq.\,\eqref{eq:Dkin_W}. What were boundary terms in the case of sharp redshift cuts now depend on the shape of the selection function $W_b(z)$ within the bin. Although technically not an exact nomenclature, we consider these to be \textit{smooth} boundary terms, and we will refer to them as such or simply as boundary terms in the following.  Again, it should be emphasized that, if $W_b(z)$ is inferred from data, then all terms in this expression are computable from observations, just as with Eq.\,\eqref{eq:Dkin_tomography}. This expression further allows us to study the expected boundary terms for various analytical and numerical choices of $W_b(z)$.  

\subsection{Examples}
\label{sec:redshiftselectionfunctions_examples}

As in Sec.~\ref{sec:theory_examples}, we study the boundary terms, now defined in terms of the selection function as
\begin{align}
    B(z_b)\equiv-\int_0^\infty{\rm d}z\,f_b(z)\frac{{\rm d}\log W_b(z)}{{\rm d}\log(1+z)}\,.\label{eq:B_Wb}
\end{align}
We first establish consistency with Sec.~\ref{sec:theory_boundaryterms} by considering for $W_b$ a simple top hat function as before, $W_b(z)=1$ for $z_1<z<z_2$ and $0$ otherwise.  A derivative of a top hat returns Dirac delta functions at its edges, with which we find
\begin{align}
    B(z_b) &= - \int_0^\infty{\rm d}z\,(1+z)f_b(z)[\delta(z_1)-\delta(z_2)]\frac{1}{W_b(z)}\label{eq:Dkin_W_tophat1}\\
    & = \left.(1+z)f_b(z)\right|^{z_2}_{z_1},\label{eq:Dkin_W_tophat2}
\end{align}
in agreement with Eq.~\eqref{eq:B_tomography}, where we assumed that the remaining boundary terms $(1+z)f_b(z)\big|_0^\infty$ vanish.

As a second example, we consider $W_b(z)$ to be Gaussian, as would be expected if $z'$ is drawn from a Gaussian probability distribution centered on $z$ with constant variance $\sigma_z^2$ and $W_b'(z')$ is a very narrow top hat selection on $z'$, centered on $z_b$.  Hence, with $W_b(z)\propto\exp(-(z-z_b)^2/2\sigma_z^2)$, we find
\begin{align}
    B(z_b) & = \sigma_z^{-2}\left[\int_0^\infty{\rm d}z\,z^2f_b(z)-\left(\int_0^\infty{\rm d}z\,z\,f_b(z)\right)^2\right].\label{eq:B_W_Gaussian}
\end{align}
If $f(z)=\hbox{const.}$ on a certain redshift range sufficiently away from $z=0$, then $f_b(z)$ becomes Gaussian.  In turn, the term in square brackets returns the redshift variance, implying $B(z_b)=\sigma_z^{-2}\sigma_z^2=1$, just as in the top panels of Fig.\,\ref{fig:boundary1}. This gives the same boundary terms as in the case where $W_b(z)$ was a top hat, as discussed below Eq.\,\eqref{eq:B_tomography}.  This is because in this situation, roughly as many observed redshifts spread to the boundary's left and right, due to the symmetry of the Gaussian uncertainties.  This preserves the amplitude of the dipole in each redshift bin, or more specifically the amplitude of the boundary terms $B$.  However, if $f(z)$ is not constant, this symmetry is broken and $W_b(z)$ becomes skewed.  In those cases, a top hat selection on $z'$ is described by the general expression Eq.~\eqref{eq:Dkin_W}.

\subsection{Flux-cut dependence}
\label{sec:redshiftselectionfunctions_s*dependence}

The kinematic dipole amplitude in redshift bins, defined via general selection functions $W_b=W_b(z)$, see Eq.~\eqref{eq:Dkin_W} and App.~\ref{app:derivation1}, is predicted entirely from quantities observed in the projected distributions. There, the prediction makes no reference to luminosity functions; just as is the case for the fully redshift-integrated dipole amplitude, Eq.~\eqref{eq:DEB}~\citep{vonHausegger:2024jan}.  This is only relevant, of course, if sufficient redshift information is available, such that the selection function $W_b(z)$ can be constructed.  For photometric redshift measurements, we must consider as a further complication the possible dependence of the redshift selection function on the flux cut, i.e.~$W_b=W_b(z;S_*)$.\footnote{Note that this is an instance separate from the flux cut dependence of the integral source counts themselves, $n(z)=n(z,S>S_*)$ which has been considered all along.}  In practice, such a dependence would arise from Eq.~\eqref{eq:Wb_main} where the conditional probability distribution $P=P(z',z;S_*)$ has been formed from two cross-matched samples $n(z,S>S_*)$ and $n'(z',S>S_*)$ each carrying a dependence on $S_*$.  Instead of considering the integral source counts, $n(z,S>S_*)$, one might also consider a flux dependence of a redshift selection function $\widetilde{W}_b(z,S)$ that acts on the differential source counts $\tilde n(z,S) \equiv \dd N/ \dd\Omega\dd z\dd S(z,S)$.  In this case, one defines
\begin{align}
\widetilde{W}_b(z,S) = \int_{0}^{\infty} \dd z'\widetilde{W}'_b(z',S) P(z',z;S)\,,
\end{align}
such that
\begin{align}
n_b(z,S>S_*) = \int_{S_*}^{\infty} \dd S\, \t{n}(z,S) \widetilde{W}_b(z,S)\,,\label{eq:differential_WzS}
\end{align}
as the analogs of \eqref{eq:Wb_main} and \eqref{eq:nb_main}, respectively.

It is precisely the distinction between $W_b(z;S_*)$ and $\widetilde{W}_b(z,S)$ that prevents a simple resolution to computing the kinematic matter dipole expectation in photometric redshift bins whose selection functions are not constant with flux $S$, and therefore not with flux cut $S_*$ either.  Instead, one may apply Eq.~\eqref{eq:Dz_int_tomography} with Eq.~\eqref{eq:Dz2} or alternatively Eq.~\eqref{eq:Dz1}, if one is ready to assume a cosmological model.  The former option requires the determination of $\alpha(z)$, $x(z)$, and $n(z)$, which may be measured from a spectroscopic subsample of sources, and $f_b(z)$, which is the most challenging step.  It requires knowledge of the differential selection function $\widetilde{W}_b(z,S)$ for each flux bin to compute $n_b(z,S>S_*)$ via Eq.~\eqref{eq:differential_WzS}.  Finally, with $n_b(z,S>S_*)$ at hand, $f_b(z)$ which enters in Eq.~\eqref{eq:Dz2}, can be computed by normalizing
\begin{align}
f_b(z) = \frac{n_b(z,S>S_*)}{\int_0^\infty \dd z \,n_b(z,S>S_*) }\,.
\end{align}
It may well be that there is a way to express the integral in Eq.~\eqref{eq:Dz_int_tomography} for the case where $\p_{S}\widetilde{W}_b(z,S)\neq 0$, but we leave this for future work when observational data will be able to provide guidance regarding the specific dependencies to-be-expected for the respective measurements.

In the following section, we consider plausible selection functions acting on redshift distributions expected in some large-scale galaxy surveys for the prediction of the boundary terms' magnitude.  Thus far, a flux cut dependence of their selection functions has not been discussed in the context of forecasts, so we too refrain from including above considerations in our forecasts that follow.  Nevertheless, we note that this dependence can lead to non-zero contributions to the expected boundary terms and therefore should not be ignored in future redshift-tomographic investigations of the kinematic matter dipole.

\section{Practical examples}
\label{sec:practicalexamples}

Large galaxy catalogs from current and near-future experiments with large sky coverages are eagerly anticipated in the context of kinematic matter dipole measurements.  Most notably, these include optical and near-IR measurements by the \textit{Euclid} satellite~\cite{2024arXiv240513491E}, mid-IR measurements by \textit{SPHEREx}~\cite{SPHEREx:2014bgr}, optical spectroscopic measurements by the Dark Energy Spectroscopic Instrument~\cite[DESI,][]{DESI:2019jxc}, optical photometric measurements in the Legacy Survey of Space \& Time (LSST) by the Vera C.~Rubin Observatory~\cite{2019ApJ...873..111I} as well as the Dark Energy Survey~\cite[DES,][]{DES:2025xii}, and the radio surveys of the Square Kilometre Array~\citep[SKA,][]{2020PASA...37....7S}.  The catalogs resulting from each of these surveys will undergo sophisticated modeling of their redshift distributions by a combination of photometric and spectroscopic techniques. 

In this section, we consider forecasted redshift distributions and uncertainties for Euclid, \textit{Rubin}-LSST and SKA. We compute their boundary term contributions to the kinematic matter dipole in tomographic redshift bins. For these redshift distributions, the trend of the boundary terms with effective redshift generically follows a monotonically decreasing function which crosses zero roughly at the peak redshift of the source distribution.  Since these three surveys collectively cover optical, near-IR, and radio frequencies, we consider their boundary terms to be sufficiently representative also for other surveys with similar sensitivities and at the same frequencies.  This is simply because the boundary terms depend mainly on the considered redshift distribution of sources and only mildly on the redshift uncertainties.  For instance, it is expected that \textit{SPHEREx} and DESI deliver catalogs with redshift distributions closely matching the one modeled for \textit{Euclid} here.  Likewise, our model of the \textit{Rubin}-LSST redshift distributions resembles those of the public DES observations. Detailed studies, that also include other survey-specific aspects which might affect number count dipole measurements such as dust extinction and the influence of spectral lines, are therefore left for future work.

\subsection{\textit{Euclid} and \textit{Rubin}-LSST}
\label{sec:practicalexamples_euclidrubin}

Following the specifications in the \textit{Euclid} Red Book~\cite{2011arXiv1110.3193L} (see also Ref.~\cite{2020A&A...642A.191E} for a more recent treatise), and those in LSST Dark Energy Science Collaboration (DESC) Science Requirement Document~\cite[SRD,][]{2018arXiv180901669T} we model the redshift distribution of both \textit{Euclid} and \textit{Rubin}-LSST by
\begin{align}
n(z) = z^2 \hbox{exp}\l[- \l(z/z_0\r)^\rho\r]\label{eq:Smail}.
\end{align}
We consider redshifts $z$, sampled from this distribution, to be observed as redshifts $z'$ that have been normally scattered around $z$ with standard deviation $\sigma_z(z) = \sigma\cdot (1+z)$.  Top hat selections on the observed redshifts $W_b'(z')$ are correspondingly convolved with the Gaussian conditional distribution
\begin{align}
    P(z',z) = \frac{1}{\sqrt{2\pi\sigma_z^2}}\exp\left[-\frac{1}{2}\left(\frac{z'-z}{\sigma_z}\right)^2\right],
    \label{eq:P_nobias}
\end{align}
to arrive at the selection functions $W_b(z)$ as per Eq.\,\eqref{eq:Wb_main}. These, in combination with $n(z)$ of Eq.\,\eqref{eq:Smail}, give the distributions of sources per bin $n_b(z)$, Eq.~\eqref{eq:nb_main}.\\

For \textit{Euclid}, the forecasted distribution parameters read $\rho=1.5$ and $z_0=0.9/\sqrt{2}\simeq 0.64$; the redshift uncertainty scales with $\sigma=0.05$. Our selections of redshifts $z'$ are chosen such that each bin carries the same number of sources. To this effect, ten bins distributed within the range $0<z'<2.5$ are bounded by the bin-edges $\{0.001, 0.42, 0.56, 0.68, 0.79, 0.9, 1.02, 1.15, 1.32, 1.58, 2.5\}$, where neighboring values are the minimum and maximum redshift $z_1'$ and $z_2'$ of the respective bin~\cite{2020A&A...642A.191E}.

\begin{figure}
    \centering
    \includegraphics[width=0.99\columnwidth]{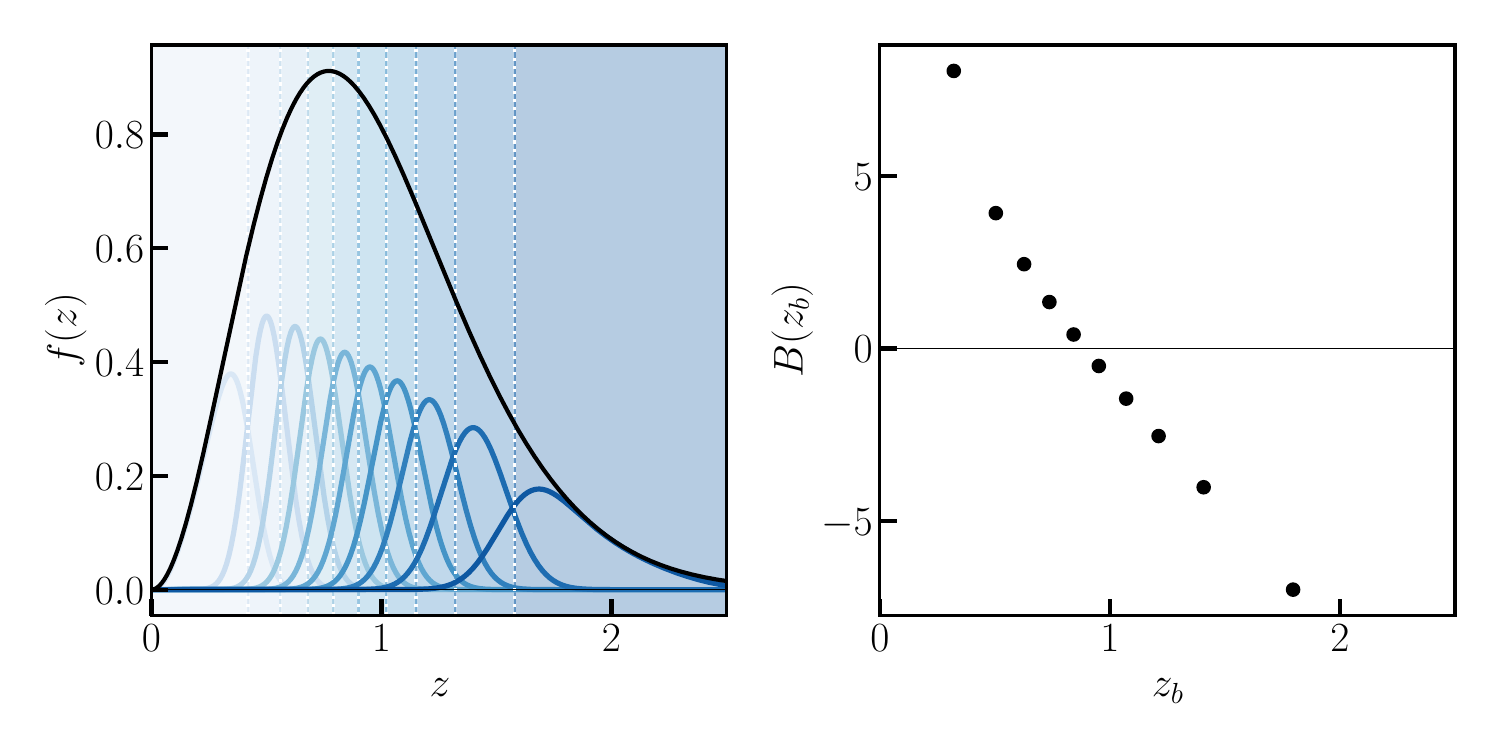}
    \caption{Computation of boundary terms, Eq.~\eqref{eq:B_Wb} for \textit{Euclid} specifications.  \textit{Left panel:} \textit{Euclid}'s $n(z)$ (black line) along with its equal-populated bins $n_b(z)$ (blue lines) selected by top hats from $z'$ (blue-shaded backgrounds) as described in the main text.  \textit{Right panel:} Boundary terms $B(z_b)$ calculated via Eq.~\eqref{eq:B_Wb} corresponding to the distributions $n_b(z)$ on the left.  These are to be compared with the $\mathcal{O}(1)$ dipole amplitudes expected from Eq.~\eqref{eq:DEB}.}
    \label{fig:Euclid_example}
\end{figure}

Fig.~\ref{fig:Euclid_example} shows \textit{Euclid}'s predicted redshift distribution $n(z)$ and its binned distributions $n_b(z)$, Eq.~\eqref{eq:nb_main}, in the left panel, where we also indicate the corresponding top hat selections by the shaded boxes. The right panel shows the bins' boundary terms $B(z_b)$ that contribute to the kinematic dipole.  Notably, the boundary terms change sign roughly at the peak of the redshift distribution and their magnitudes are comparable to (or even in excess of) the $\mathcal{O}(1)$ dipole amplitude from Eq.~\eqref{eq:DEB}.\\

For \textit{Rubin}-LSST, we follow both the year 1 (Y1) and year 10 (Y10) targets for the LSS (large scale structure) clustering analysis\footnote{See Appendix D1 of the SRD~\cite{2018arXiv180901669T}.  Instead of the lens sample, we could have also chosen the source sample whose parameters are found in Appendix D2.  However, for the purpose of this work, this distinction is not significant.} by setting $\sigma=0.03$.  The forecasted redshift distribution parameters are $(\rho,z_0)=(0.94,0.26)$ and $(0.90,0.28)$, respectively.  Here, the flux limits of the catalogs are provided by $i$-band magnitude limits of $i_{Y1}=24.1$ and $i_{Y10}=25.3$.  To stay in line with the Science Requirements Document (SRD), we bin redshifts $z'$ by 5(10) equal-width top hats, arranged contiguously between $z'=0.2$ and $z'=1.2$ to simulate the Y1(Y10) targets.

\begin{figure}
    \centering
    \includegraphics[width=0.99\columnwidth]{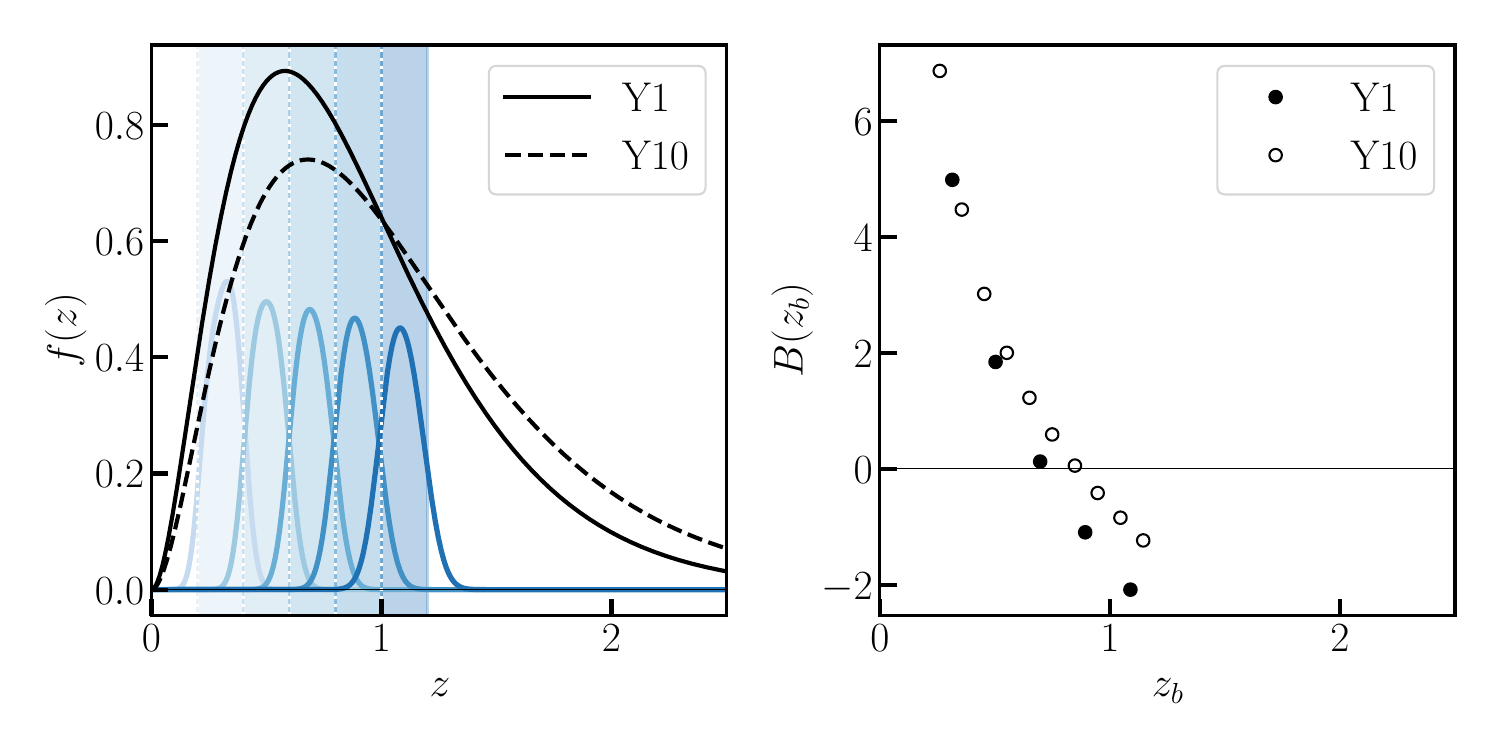}
    \caption{Computation of boundary terms, Eq.~\eqref{eq:B_Wb}.  Like Fig.~\ref{fig:Euclid_example} but for \textit{Rubin}-LSST Y1(Y10) specifications, shown as a solid (dashed) black line in the left panel and filled (open) circles in the right panel.  In both cases, these boundary terms can outweigh the dipole contributions from Eq.~\eqref{eq:DEB}, especially at low redshift.}
    \label{fig:LSST_example}
\end{figure}

Fig.~\ref{fig:LSST_example} shows the predicted redshift distributions $n(z)$ for the Y1 and Y10 targets.  For visual clarity, we only show the Y1 binned distributions $n_b(z)$, Eq.~\eqref{eq:nb_main}, and again indicate the equi-distant equal-width bins  by shaded boxes.  The right panels show the bins' boundary terms $B(z_b)$ highlighting the dependence of boundary terms on the source distribution, even for the same experiment.  As for Euclid, these boundary terms may well outweigh the standard dipole contributions from Eq.\,\eqref{eq:DEB}.\\

Both experiments' boundary terms exhibit similar behavior, crossing over from a positive to a negative contribution to the kinematic matter dipole amplitude close to the maximum of their redshift distribution, as expected from Eq.~\eqref{eq:Dkin_W_tophat2}.  The magnitude of $B(z_b)$ is below unity only in this cross-over region, where the kinematic dipole amplitude would be dominated by the first terms in Eq.~\eqref{eq:Dkin_W} and hence by the aberration and boosting of flux values, as opposed to the boosting of redshifts.

As remarked above, the quoted redshift distributions $n(z)$ correspond to fixed flux limits. It is of course possible to perform the \citeauthor{1984MNRAS.206..377E} test for different flux limits within the survey sensitivity, which would change $\t{x}$ and $\t{\alpha}$ on top of the models' defining parameters $\rho$ and $z_0$ and therefore also the shape of the distribution Eq.~\eqref{eq:Smail}, which more generally is determined by the luminosity function. This evolution is included in the first line of Eq.~\eqref{eq:Dkin_W}~\cite{vonHausegger:2024jan}, provided that $\tilde x$ and $\tilde \alpha$ are computed anew for and at this new flux cut.  New in this context is that it also affects the last term and so also the boundary terms, Eq.~\eqref{eq:B_Wb}.  We can see a hint of this in Fig.~\ref{fig:LSST_example}, where the Y1 and Y10 distributions lead to boundary terms that each lie on slightly different curves in the right panel.\footnote{In fact, the curve on which the boundary terms appear to lie, i.e.~also their magnitude, is dominated by the shape of the redshift distribution $n(z)$ rather than the bin selection parameters $\sigma_z$ or the bin spacing, especially for narrow bins.  We offer an explanation for this in Appendix~\ref{app:moreparameters_bin}.}

\subsection{SKA}
\label{sec:practicalexamples_ska}

\begin{figure}
    \centering
    \includegraphics[width=0.99\columnwidth]{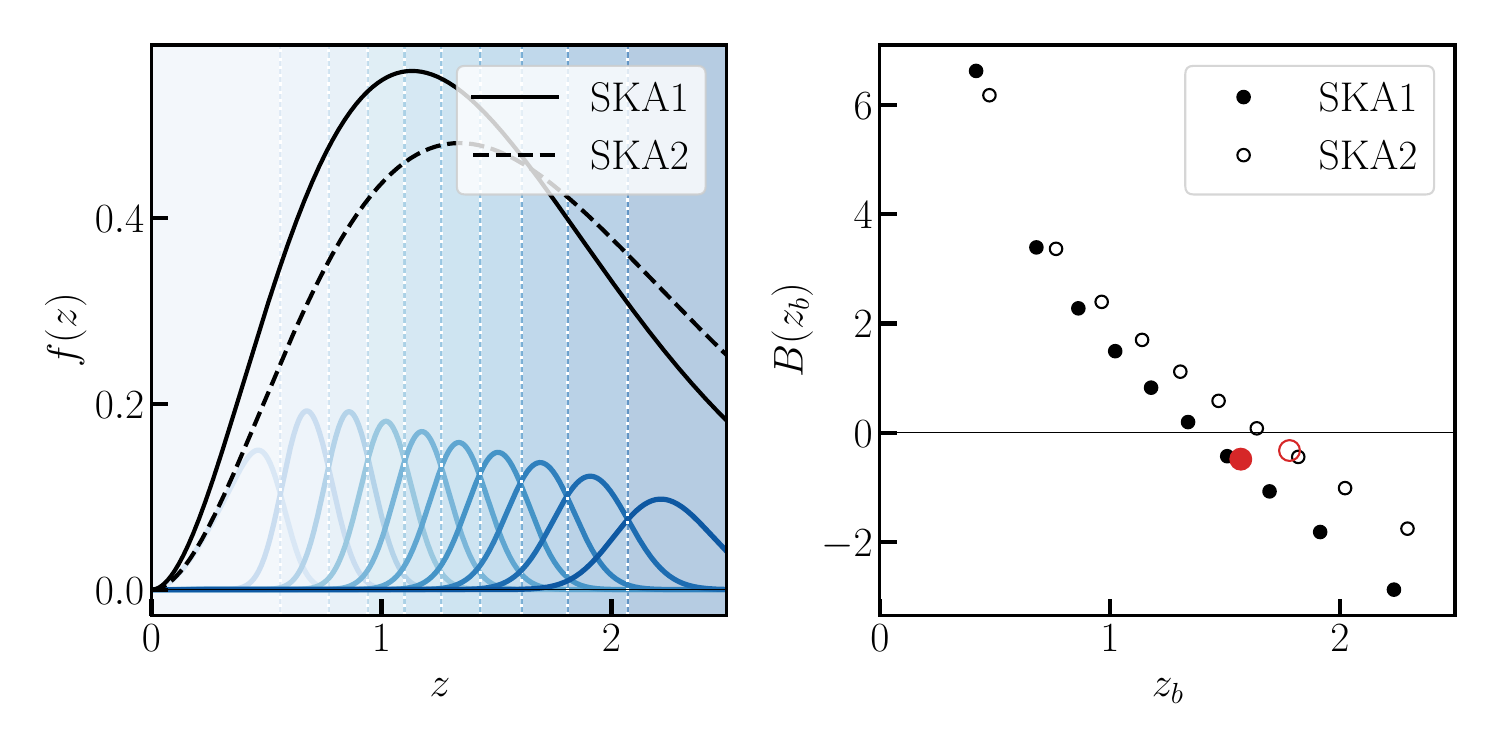}
    \caption{Computation of boundary terms, Eq.~\eqref{eq:B_Wb}.  Like Fig.~\ref{fig:LSST_example} but for SKA1 and SKA2 specifications.  The right panel also includes as red points the boundary terms expected for the subsample $z>0.5$ as suggested to remove `local structure'.  Also here, the boundary terms can exceed the dipole contribution from Eq.~\eqref{eq:DEB}.}
    \label{fig:SKA_example}
\end{figure}

With flux sensitivities as deep as $\sim\mu{\rm Jy}$, galaxy catalogs expected from the SKA will contain a range of different source types, among others, star forming galaxies (SFG) and AGN.  Detailed studies~\cite{2008MNRAS.388.1335W,2015ApJ...814..145A} of the corresponding luminosity functions have delivered predictions of redshift distributions of the various source types at different flux limits.  However, to enable comparison with the previous subsection's results, we adopt the model parameters of Ref.~\cite{2016MNRAS.463.3674H} for the same distribution as above, Eq.~\eqref{eq:Smail}, and for predominantly SFGs, for which $(\rho,z_0)=(1.25,1.1/\sqrt{2}\simeq 0.78)$ and $(1.25,1.3/\sqrt{2}\simeq 0.92)$, respectively. The redshift uncertainties, that scale with $z$ as above, are set to $\sigma=0.05$ and $0.03$.  Details on the derivation of these parameters are found in Ref~\cite{2016MNRAS.463.3686B}; we here simply present the two configurations as SKA1 and SKA2, both of which achieve $\sim\!\mu{\rm Jy}$ flux limits and include source sizes down to $\sim\!0.5''$.  Similar to the choice above for Euclid, we pick equal-populated redshift bins for each of SKA1 and SKA2 with the bin edges $\{0.001, 0.56, 0.77, 0.94, 1.1, 1.26, 1.43, 1.61, 1.81, 2.07, 2.5\}$, $\{0.001, 0.64, 0.86, 1.05, 1.22, 1.39, 1.56, 1.73, 1.93, 2.15, 2.5\}$, respectively.  We show the corresponding redshift distributions and boundary terms in Fig.~\ref{fig:SKA_example} as before. Again, for visual clarity, we only show the binned distributions for SKA1.\\

The possibility for SKA to conclusively measure the matter dipole at radio frequencies, even for different source types such as AGN and SFGs, is enticing.\footnote{The first measurements of the matter dipole with predominantly radio SFGs was performed in Ref.~\cite{2024arXiv240816619W} where just this motivation is highlighted.  However, even relatively few sources remaining in their final selection prevent a conclusion as to whether the matter dipole amplitude is consistent or inconsistent with the CMB dipole prediction, requiring larger surveys such as the SKA to settle this question for SFGs.}  However, whichever source type is cataloged, it is expected that the boundary terms will remain in a similar range -- comparable in size to the dipole contributions of aberration and flux boosting and therefore non-negligible for a robust interpretation of observed matter dipole amplitudes.  Particularly noteworthy in the present context of SKA are the growing boundary terms towards smaller redshifts:  The matter dipole contribution by local structure has been the focus of many a study of systematic issues resulting in the general understanding that the removal of low-redshift sources is essential for a clean measurement of the kinematic matter dipole. This is simply due to the large dipole contribution expected from the clustering of sources on smaller scales.  It has been suggested to remove all sources with photometric redshifts $z'<0.5$ to ensure a sample that is virtually unaffected by the truly clustered sources at $z<0.1$~\cite{2019MNRAS.486.1350B}. This is certainly a good idea, although it has not been investigated what effect this removal would have on the measured dipole amplitudes.\footnote{While we do not dispute the conclusion of Ref.~\cite{2019MNRAS.486.1350B} that the clustering of sources with $z<0.5$ contributes uncertainty to the sought-after kinematic matter dipole, we would like to point out that the application of a strict cut in redshift causes a kinematic dipole amplitude that is in fact guaranteed to be affected by the boundary terms discussed here.}  Strictly \textit{removing} sources with $z<0.5$ for SKA1(SKA2), we find the expected boundary terms to be $-0.48$($-0.33$), Eq.~\eqref{eq:B_tomography}, so leading to a kinematic matter dipole expectation relatively smaller than the Ellis\&Baldwin results by $\mathcal{O}(10\%)$ for typical radio parameters.  We show these values among the other boundary terms in the right panel of Fig.~\ref{fig:SKA_example} at their respective effective redshift $z_b$.  On the other hand, the kinematic dipole expectation for the sources \textit{selected} by $z<0.5$, would be increased by boundary terms of $+7.55$($+7.82$), Eq.~\eqref{eq:B_tomography}, for SKA1(SKA2), or a $\mathcal{O}(100\%)$ relative increase, even without the clustering contribution.

\subsection{Biased redshift estimates}
\label{sec:practicalexamples_biasedredshifts}

In many ways, the results presented here represent best-case scenarios in which redshifts $z'$ are scattered around $z$ without any bias or unexpectedly large variance. Works such as Refs.~\cite{2020A&A...642A.191E,2024arXiv240917377J,2024arXiv240914265A} have also considered the influence of biased redshift estimates as well as that of catastrophic outliers.\footnote{Catastrophic outliers refer to a subpopulation of sources for which the inferred photometric redshift is grossly biased compared with the remaining redshift estimates.}  Naturally, even in such cases, if the bin selection function $W_b(z)$ is perfectly known, the estimated boundary terms do not deviate from those reported here.  However, if $W_b(z)$ itself is misestimated, then the boundary terms are also biased.  We explore some examples of this in App.~\ref{app:moreparameters_bias}, where we also propose an approximate solution in the absence of knowing the selection function perfectly.  Here, to provide an example, we consider the \textit{Euclid} redshift distribution with all parameters fixed as previously in Sec.~\ref{sec:practicalexamples_euclidrubin}, however, where $10\%$ of the sources have photometric redshift estimates biased additively by $0.1$ (we replace Eq.~\eqref{eq:P_nobias} with Eq.~\eqref{eq:P_bias} and set $f_o=0.1$, $a_o=0.1$, while keeping $a=0$, $m=m_o=1$, and $\sigma_z=\sigma_{z,o}$, corresponding to the case in Table 5 of Ref.~\cite{2020A&A...642A.191E}).  While generally the accurate modeling of redshift distributions and corresponding redshift bin selections is important for the computation of accurate kinematic dipole expectations, the presented example of moderate bias does not pose serious deviations from the boundary terms computed if biases were ignored ($\mathcal{O}(0.1\beta)$ corrections to $\mathcal{O}(4\beta)$ dipole amplitudes). Nevertheless, as an exercise, we provide other examples in App.~\ref{app:moreparameters_bias}.

\begin{figure}
    \centering
    \includegraphics[width=0.99\columnwidth]{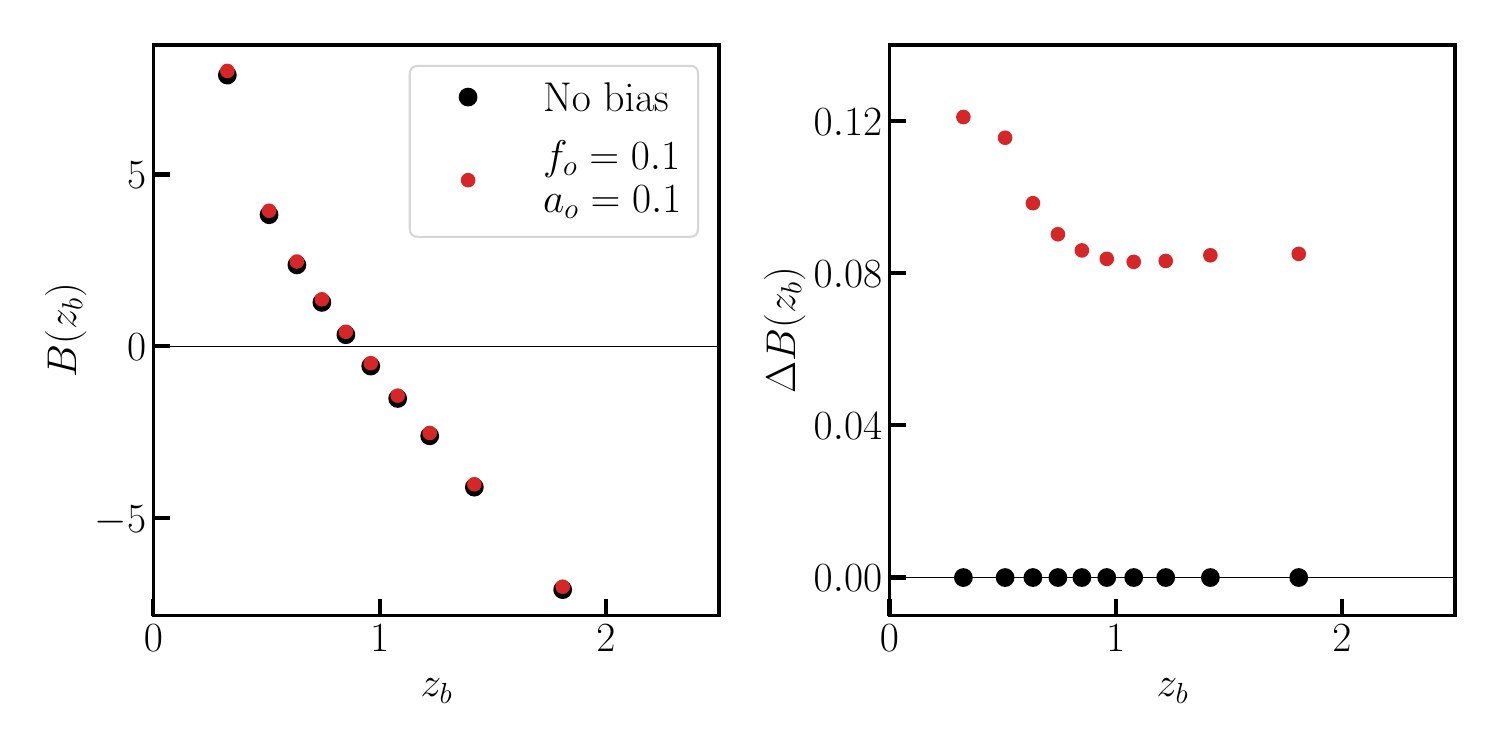}
    \caption{Boundary terms including redshift bias for a $10\%$ fraction of `outliers' using \textit{Euclid} specifications.  $B(z_b)$ computed  with Eq.~\eqref{eq:B_Wb} and Eq.~\eqref{eq:P_nobias} (black dots) are compared with those computed with Eq.~\eqref{eq:B_Wb} and Eq.~\eqref{eq:P_bias} (red dots) where the fraction of outliers is $f_o=0.1$ and their additive bias to $z$ is $a_o=0.1$.  \textit{Left panel:} Boundary terms.  \textit{Right panel:} Difference of boundary terms $B_{\rm bias}-B_{\rm no bias}$.  This example leads to only negligible corrections to the typical dipole contributions of Eq.~\eqref{eq:DEB}.}
    \label{fig:Euclid_biasexample}
\end{figure}

\section{Discussion}
\label{sec:discussion}

The recent interest in studying kinematic observer effects on galaxy surveys and CMB alike has resulted in the creation of a number of tests, beyond just the \citet{1984MNRAS.206..377E} dipole.  We therefore would like to offer a brief distinction between the effects discussed here and some of those others in the recent literature that too consider redshift tomography.

Our work concerns redshift-tomographic measurements of the dipole in galaxy number counts.  The definition of bins in observed redshift brought about additional, so far unaddressed contributions to the number count dipole, simply due to the boost to background redshifts, via Eq.\,\eqref{eq:z_boost}. From the same equation it is evident that the observed redshifts themselves carry a dipolar signal which can be studied independently of tracing the galaxies' number counts.  Indeed Refs.~\cite{2024arXiv240314580D,2024arXiv240909946T} do just this, by using spectroscopic redshift measurements and by binning in a $\beta$-dependent `reconstructed' background redshift $\bar{z}$ in order to fit for the so-called redshift dipole.\footnote{Instead binning in \textit{observed} redshift $z$ would not give a measurable effect.}  It should be noted that tomographic measurements were rather a means to an end for attempting to detect the redshift dipole, not an essential feature of the phenomenon.  In contrast, redshift selections \textit{are} an essential step in measuring the redshift dependence of the kinematic number count dipole, and the boundary terms discussed here are an immediate result thereof.

Further, it is of course possible to define estimators that would include both, a contribution of the redshift dipole \textit{and} those inherent to the number count dipoles, by weighting galaxy number counts with redshifts -- or any boost-affected quantity for that matter.  Redshift-weighted number counts were so far only discussed in the context of measuring the galaxies' peculiar velocities on small scales, in order to measure the cosmic growth factor and other quantities associated with redshift-space distortions~\citep[e.g.][]{Hernandez-Monteagudo:2019epd,Legrand:2020sek,Matthewson:2022bpp}.  An effect not discussed in that context is our observer velocity with respect to the galaxies, although corresponding modifications to include our observer boost in the observables might be possible. To this effect, other weights to galaxy number counts than redshifts had been considered earlier in Ref.~\cite{2011ApJ...742L..23S} who used the observed flux densities as weights per galaxy.  Recently this idea was generalized by Ref.~\cite{Nadolny:2021hti} to arbitrary weights composed of redshift, size, and flux density, in order to find those weights that optimally distinguish kinematic from non-kinematic contributions to the number count dipole.  With reference to our present work, it is worthwhile to compute the boundary terms expected also in such weighted estimators if applied to tomographic subsamples.\\

The boundary terms that appear for measurements in redshift bins add to the Ellis \& Baldwin dipole amplitude, Eq.~\eqref{eq:DEB}, thereby altering the kinematic dipole expectation for a given value of $\beta$.  That said, we note that the boundary terms too are proportional to $\beta$ and therefore may assist a measurement of the kinematic observer boost, rather than complicate it.  Just like in the case without boundary terms, however, the boost is measured with respect to the sample of sources under investigation, such that the sources' individual motion can in principle contribute to the measured dipole as well.  In isotropic cosmologies, these individual contributions are expected to average out.  However, if the sources were part of a coherent large-scale bulk flow~\citep[see e.g.][]{Watkins:2023rll,Whitford:2023oww}, one would find also the boundary terms to be enhanced, by the corresponding enhancement of $\beta$.

As we have seen, the boundary terms themselves can outweigh those terms known from Eq.\,\eqref{eq:DEB}, by a factor of a few or more, depending on the redshift distribution of the sample. For this reason, the presence of $\beta$-dependent boundary terms might increase the prospect of measuring the kinematic dipole signal in galaxy number counts for samples that otherwise did not have sufficient size and therefore too low signal-to-noise. As we reach higher precision on the kinematically generated dipole, one may wonder if the kinematically generated quadrupole, which is expected to be generated at second order in $\beta$ would become detectable as well -- this would offer yet another test of the kinematic nature of any measured matter dipole. While the expected quadrupole would be down-weighted by $\beta^2\sim\mathcal{O}(10^{-6})$, the corresponding pre-factor (analogous to that in Eq.~\eqref{eq:Dkin_W}) might elevate the quadrupole's amplitude beyond the $\mathcal{O}(10^{-5})$ quadrupole given by local clustering of structure.  While near-future surveys are expected to deliver enough sources to significantly suppress shot noise, a thorough forecast of their ability to detect such second-order effects would require quantification of the noise on the boundary terms as well; we leave this for future work.\\

In the search for finding the origin of so-far reported excess number count dipoles, tomography in individual redshift bins will certainly prove itself invaluable. The present work paves the way to perform the corresponding interpretation of such measurements correctly.  Moreover, this work opens the door to a number of other tests with redshift-tomographic measurements.  These include cross-bin analyses of the kinematic number count dipole, where neighboring redshift bins will exhibit the strongest correlation, owing to the defining nature of the discussed boundary terms -- namely, that sources are boosted out of or into the considered redshift bin, into or from the respective neighboring bin.  Simultaneously, cross-bin-correlation of the noise would be expected small but non-zero if the  uncertainties on the redshift distribution estimates correlate too.  Going forward, this test will also allow for redshift-tomographic cross-correlation between galaxy number counts and other tracers, such as weak lensing, which too might carry signatures of anomalous clustering.\\

So far observational studies of the \citeauthor{1984MNRAS.206..377E} test have not explicitly restricted their sample in redshift, largely due to the absence of corresponding data.  However, cuts in other variables may have been applied as part of the sample selection, as for instance the color cuts in Ref.~\cite{Secrest:2020has} (see also~\cite{Panwar:2024xum}), or even inadvertently as part of the survey specifications. In light of our present work, it is essential to understand which cuts may imply selections on redshift or on any other quantity affected by Doppler boosts or aberration, simply because of their correlation. This may in turn lead to boundary terms \textit{à la} Eq.~\eqref{eq:B_Wb}. We briefly illustrate this point using the example of a color cut and refer to App.~\ref{app:otherselection} for more details.

Consider a strict color cut $C>C_*$, applied to a sample of sources,
\begin{align}
    n_C(z,S_*,C_*) =& \int_{C_*}^\infty{\rm d}C\, \tilde n_C(z,S_*,C),
\end{align}
where we defined the source counts, integrated over flux \textit{and} color, as \mbox{$n_C(z,S_*,C_*) = {\rm d}N/{\rm d}\Omega{\rm d}z(z,S>S_*,C>C_*)$,} which are related to their differential counterpart $\tilde n_C(z,S_*,C) = {\rm d}N/{\rm d}\Omega{\rm d}z{\rm d}C(z,S_*,C)$.  Color $C$ is commonly correlated with \textit{observed} redshift via the $k$-correction. Hence, one way to model this correlation might be via a conditional distribution $P_C(C,z)$ in analogy to the function $P(z',z)$ of Eq.~\eqref{eq:nprime_convolution_main}.  We then find
\begin{align}
    n_C(z,S_*,C_*) \simeq &\int_{C_*}^\infty{\rm d}C\, P(C,z) \tilde n(z,S_*)\label{eq:nC_split}\\
    \equiv& \,n(z,S_*)W_C(z)\,.
\end{align}
where we defined a redshift selection function $W_C(z)$ just as in Eq.~\eqref{eq:Wb_main}. Analogously to our Sec.~\ref{sec:redshiftselectionfunctions_}, we see that a strict cut on color, e.g.~$C>C_*$, implies the presence of \textit{smooth} boundary terms Eq.\,\eqref{eq:B_Wb}, even in the absence of explicit redshift information. It is important to note, however, that our ansatz, Eq.~\eqref{eq:nC_split} is not always valid, as $P_C(C,z)$ may also depend on flux.  We encountered this issue already in Sec.~\ref{sec:redshiftselectionfunctions_s*dependence} and hence it is more principled to perturb the full distribution $n_C(z,S_*,C_*)$ in terms of its variables as, e.g., done in Ref.~\cite{Dalang:2021ruy}, but here including color.  This alludes to the understanding that, a color cut $C>C_*$ may sample the integral source counts at different redshifts in the forward and backward directions.  We expand on this point in App.\,\ref{app:otherselection_color}, where we demonstrate that for perfect power law spectra $S(\nu)\propto\nu^{-\alpha}$, Eq.~\eqref{eq:DEB} holds after all, where the redshift-dependence of color cancels, and hence $C$ does not depend on the motion of the observer.  However, deviations from power law spectra, here modeled by a running of the spectral index $\alpha=\alpha(\nu)$ may generate an additional contribution to the Ellis and Baldwin formula.  We detail this second point in App.\,\ref{app:otherselection_running_spectral_index}.\\

Lastly, in light of the vast amounts of upcoming data from large-scale galaxy surveys, we urge the inclusion of the effect the boundary terms have on the kinematic number count dipoles in corresponding forecasts and simulations.  Few works to-date have explicitly considered redshift-dependent measurements of the kinematic number count dipoles.  However, those that did~\citep{2019MNRAS.486.1350B,2023JApA...44...22G} simply assumed the validity of Eq.~\eqref{eq:DEB} per redshift bin, which was subsequently fixed as a dipolar modulation in their mock data, as opposed to discovering the boundary terms discussed here via explicit boosting of mock redshifts prior to bin selection.  In this context, if one wants to separate the respective contributions of the Ellis \& Baldwin and the boundary terms, it might be of interest to vary the flux density cut $S_*$ repeatedly, to change their relative contribution to the overall dipole amplitude.  Future redshift data will offer ample opportunity to study how these effects are imprinted onto real galaxy samples.

\section{Conclusion}
\label{sec:conclusion}

Measurements of the cosmic matter dipole offer a foundational consistency check of the standard cosmological model, by enabling direct comparison between the rest frames of matter and radiation.  In this work, we considered the measurement of the kinematic number count dipole in redshift bins. We showed that boosted redshifts generate non-negligible contributions to the dipole amplitude at the edges of the bins in redshift, adding to the expectation of the kinematic matter dipole of~\citet{1984MNRAS.206..377E}.  The latter remains accurate for number count distributions that vanish at the redshift selection boundaries, while the additional contributions, dubbed boundary terms here, can be computed with knowledge of the redshift distribution and particular redshift selection functions. These boundary terms are essential for an accurate prediction of the kinematic matter dipole in redshift-tomographic measurements and further can change with redshift bin.  In practice, one ought to measure $\tilde x$ and $\tilde\alpha$ per redshift bin with the associated sources, to obtain the Ellis \& Baldwin contribution, \textit{and} additionally compute the boundary terms per redshift bin as dictated by Eq.~\eqref{eq:Dkin_W}.  This retains the feature of kinematic matter dipole measurements to solely depend on observed sample properties without explicit reference to underlying luminosity functions.\\

We considered redshift selections using spectroscopic as well as photometric redshift, and demonstrated the effectiveness of our formalism for each. In practical examples of forecast redshift distributions and tomographic selections for \textit{Euclid}, \textit{Rubin}-LSST, and SKA, we estimated the expected magnitude of the boundary terms, which, depending on the positioning of the redshift bin, may easily exceed the Ellis \& Baldwin contribution, and may even flip the final kinematic dipole amplitude's sign (i.e.~anti-align the kinematic number count dipole with the direction of motion). If one was tempted to only consider those redshift bins with small associated boundary terms, we emphasize that the boundary terms too are proportional to $\beta$, such that their inclusion may actually assist with a measurement, rather than complicate it. Simultaneously, this places great importance on accurate estimation of redshift distributions and selection functions, which may be challenging. Nevertheless, we consider the challenges uncovered by our work to be outweighed by the many opportunities it presents.

\acknowledgments

We thank Chris Clarkson for remaining insistent on the question that sparked this work.  SvH thanks Tom Cornish and Ian Harrison for discussions on the redshift distributions of upcoming experiments.  Finally, it is a pleasure to thank Ruth Durrer and Subir Sarkar for detailed comments and helpful discussions on this manuscript.  C.D.\,is supported by the ERC Starting Grant SHADE (grant no.\,StG 949572) and thanks SPCS for hospitality.

This work made use of the following python packages: \texttt{numpy}~\citep{harris2020array}, \texttt{scipy}~\citep{2020SciPy-NMeth}, \texttt{matplotlib}~\citep{Hunter:2007}, and \texttt{healpy}~\citep{Zonca:2019vzt}.

For the purpose of open access, the authors have applied a Creative Commons Attribution (CC BY) licence to any Author Accepted Manuscript version arising

\appendix

\section{Selection functions for redshift uncertainties}
\label{app:derivation2}

This section briefly summarizes the definition of the selection functions discussed in Sec.~\ref{sec:redshiftselectionfunctions}.

Selecting a bin $b$ in observed, uncertain redshift $z'$ via a selection function $W'$ can be written as
\begin{align}
    n'_b(z') &= n'(z')W'_b(z') = \\
    &= \int_0^\infty{\rm d}z\,n(z)P(z',z)W'_b(z').
\end{align}
Number conservation $\int{\rm d}z'\,n'_b(z') = \int{\rm d}z\,n_b(z)$ leads to
\begin{align}
    &\int_0^\infty{\rm d}z' n'_b(z') \\
    =&\int_0^\infty{\rm d}z'\int_0^\infty{\rm d}z\,n(z)P(z',z)W'_b(z')\\
    =& \int_0^\infty{\rm d}z\,n(z)\int_0^\infty{\rm d}z'\,P(z',z)W'_b(z')\\
    \equiv& \int_0^\infty{\rm d}z\,n(z)W_b(z) = \int_0^\infty{\rm d}z\,n_b(z)
\end{align}
where we defined
\begin{align}
    W_b(z) \equiv \int_0^\infty{\rm d}z'\,W'_b(z')P(z',z),
\end{align}
and identified
\begin{align}
    n_b(z) \equiv n(z)W_b(z).
\end{align}

\section{Simplifying the redshift dipole}
\label{app:derivation1}

In this section, we recapitulate the essential arguments of Ref.~\cite{vonHausegger:2024jan} in the context of general selection functions $W(z)$, discussed in Sec.~\ref{sec:redshiftselectionfunctions}.  With the specification of Eqs.~(\ref{eq:Dkin_W_tophat1}-\ref{eq:Dkin_W_tophat2}) the following derivation also applies to Sec.~\ref{sec:theory_boundaryterms}.

We begin by computing the projected integral source counts of a selected redshift bin
\begin{align}
    \mathcal{N}_b(S_*) &= \int_0^\infty{\rm d}z\,n_b(z,S_*)\label{eq:ntilde1}\\
    &= \int_0^\infty{\rm d}z\,n(z,S_*)W_b(z),\label{eq:ntilde2}
\end{align}
where we re-introduced the $S_*$-dependencies, e.g.~$n(z)=n(z,S_*)$, for clarity, and where we used Eq.~\eqref{eq:nb_main}.  While the magnification bias $x(z)$ is still defined as in Eq.~\eqref{eq:x}, the local power law slope $\tilde{x}(S_*)$ of the projected integral source counts of the bin is
\begin{align}
    \tilde{x}_b(S_*) \equiv -\frac{\partial \log \mathcal{N}_b(S_*)}{\partial \log S_*}.
\end{align}
With Eqs.~(\ref{eq:ntilde1}-\ref{eq:ntilde2}) and using Eq.~\eqref{eq:x} this becomes
\begin{align}
    \tilde{x}_b(S_*) &= -\frac{1}{\mathcal{N}_b(S_*)}\int_0^\infty{\rm d}z\,\left[\frac{\partial n(z,S_*)}{\partial \log S_*}\right]W_b(z)\\
    &= \frac{\int_0^\infty{\rm d}z\,x(z)n(z,S_*)W_b(z)}{\int_0^\infty{\rm d}z\,n(z,S_*)W_b(z)}\\
    &= \int_0^\infty{\rm d}z\,x(z)f_b(z,S_*).
\end{align}

The sources contained in bin $b$ that contribute to the number count dipole are only those that lie in the vicinity of the flux density threshold $S_*$.  Hence, the relevant average of all sources' spectral indices is that of only those very sources.  To perform this average we consider the differential source counts $\tilde n$, where
\begin{align}
    n_b(z,S_*) = \int_{S_*}^\infty{\rm d}S\,\tilde n_b(z,S),\label{eq:nb_int}
\end{align}
and define an appropriate normalised distribution function
\begin{align}
    \tilde f_b(z,S_*) \equiv \frac{\tilde n_b(z,S_*)}{\int_0^\infty{\rm d}z\,\tilde n_b(z,S_*)}.\label{eq:tildef}
\end{align}
Also the differential source counts are windowed by selection function $W_b$ as $\tilde n_b(z,S_*)=\tilde n(z,S_*)W_b(z)$.  By Leibniz' rule we rewrite Eq.~\eqref{eq:nb_int} as
\begin{align}
    \tilde n_b(z,S_*) &= -\frac{{\rm d}n_b(z,S_*)}{{\rm d}S_*} = -\left[\frac{\partial n(z,S_*)}{\partial S_*}\right]W_b(z)\\
    &= -\frac{n(z,S_*)}{S_*}\left[\frac{\partial \log n(z,S_*)}{\partial \log S_*}\right]W_b(z)\\
    &= \frac{1}{S_*}x(z)n(z,S_*)W_b(z)
\end{align}
where the total $S_*$-derivative reduces to a partial derivative, and we again used Eqs.~\eqref{eq:nb_main} and~\eqref{eq:x}.  With this expression, the distribution $\tilde f$ reads
\begin{align}
    \tilde f_b(z,S_*) &= \frac{x(z)n(z,S_*)W_b(z)}{\int_0^\infty{\rm d}z\,x(z)n(z,S_*)W_b(z)}.
\end{align}
The average spectral index of those sources in near vicinity of $S_*$ is defined as
\begin{align}
    \tilde \alpha_b(S_*) \equiv& \int_0^\infty{\rm d}z\,\alpha(z)\tilde f_b(z,S_*)\\
    =& \frac{\int_0^\infty{\rm d}z\,x(z)\alpha(z)n(z,S_*)W_b(z)}{\int_0^\infty{\rm d}z\,x(z)n(z,S_*)W_b(z)}.
\end{align}
Finally, we turn to the kinematic dipole amplitude, Eq.~\eqref{eq:Dz2}, averaged over redshifts, Eq.~\eqref{eq:Dz_int_tomography}, which becomes
\begin{align}
    \mathcal{D}_{\rm kin}(z_b) =& \!\int_0^\infty\!\!\!{\rm d}z\,f_b(z)\left[3+x(z)\!\left(1+\alpha(z)\right)+\frac{{\rm d}\log n(z)}{{\rm d}\log (1+z)}\right]\beta \nonumber\\
    =& \left[3+\tilde x_b(1+\tilde \alpha_b)\right]\beta + \int_0^\infty{\rm d}z\,f_b(z)\frac{{\rm d}\log n(z)}{{\rm d}\log (1+z)}\beta.
\end{align}
The remaining integral reduces as follows, again partially integrating, as before in Eq.~\eqref{eq:int_part2_eval},
\begin{align}
    &\int_0^\infty{\rm d}z\,f_b(z)\frac{{\rm d}\log n(z)}{{\rm d}\log (1+z)}\\
    =&\frac{1}{\int_0^\infty{\rm d}z\,n(z)W_b(z)}\int_0^\infty{\rm d}z\,(1+z)\frac{{\rm d} n(z)}{{\rm d} z)}W_b(z)\\
    =&\frac{1}{\int_0^\infty{\rm d}z\,n(z)W_b(z)}\\
    &\quad\times\left[(1+z)n(z)W_b(z)\big|_0^\infty-\int_0^\infty{\rm d}z\,n(z)W_b(z)\frac{{\rm d}\log W_b(z)}{{\rm d}\log(1+z)}\right] \nonumber\\
    =&\,-\int_0^\infty{\rm d}z\,f_b(z)\frac{{\rm d}\log W_b(z)}{{\rm d}\log(1+z)}\,.
\end{align}
This concludes the derivation of Eq.~\eqref{eq:Dkin_W} and generalises the correspondence made in Ref.~\cite{vonHausegger:2024jan} to redshift bins.  Similar observations as those made below Eq.~\eqref{eq:Dkin_tomography} also hold here.

\section{Additional parameter studies for \textit{Euclid} and \textit{Rubin}-LSST}
\label{app:moreparameters}

In this appendix, we discuss the approximate independence of the boundary terms $B(z_b)$ on the bin edges which preserve the average redshift of the bin $z_b$.  We then detail investigations on biased photometric redshifts, which affect the selection functions $W_b(z)$ appearing in Eq.~\eqref{eq:Dkin_W}. Lastly, we provide helpful relations, to efficiently sample redshifts numerically according to the distribution considered in this work.

\subsection{Sensitivity of boundary terms to bin choices}
\label{app:moreparameters_bin}

In section~\ref{sec:theory_examples}, boundary terms were shown for simple example distributions and top-hat selections, where the boundary terms were plotted as functions of the bins' effective redshift, $z_b$. Generally, however, boundary terms are functions of the bin edges $z_1$ and $z_2$. Yet, under certain conditions, a presentation of the boundary terms as functions of $z_b$ is a good approximation, at least for the purpose of illustration.  This can be illustrated by charting the magnitude of boundary terms and the corresponding $z_b$ for an array of choices of $z_1$ and $z_2$. Depending on the particular choice of bins the 2-dimensional functions $B(z_1,z_2;f)$ and $z_b(z_1,z_2;f)$ are sampled at the corresponding coordinates $(z_1,z_2)$.

In such a chart, equi-distant, equal-width bins are determined by $z_2-z_1=\hbox{const.}$ -- diagonal lines through the ($z_1$,$z_2$) space.  Narrower bins lie closer to the asymptote $z_2-z_1=0$, and therefore sample these 2-dimensional functions at points that lie close to one another. Then, for sufficiently smooth redshift distributions $n(z)$, variations in these diagonal lines do not pick up vast differences in the corresponding $B$ or $z_b$, justifying the approximation of the boundary terms as a function of $z_b$. Changing the bins' distance, at fixed width, simply samples the corresponding diagonal at different points, again allowing for the same approximation. However, changing the bins' widths will sample the $B(z_1,z_2)$ and $z_b(z_1,z_2)$ at points that deviate from the diagonal.  Then, the recovered boundary terms can appear to follow a different functional form, and the approximate relationship $B(z_1,z_2;b)\simeq B(z_b;f)$ breaks down.  This also occurs if bins are chosen to be comparably broad.

Where bins are chosen in uncertain, observed redshift $z'$, the situation only changes slightly. Here, the corresponding, effective bin-width in $z$ is determined also by the uncertainty $\sigma_z$.  In Fig.~\ref{fig:zb_example_2}, we show an example of $B(z_1',z_2')$ and $z_b(z_1',z_2')$ for the choice of the Euclid redshift distribution $n(z)$, where we compare the trace in $(z_1',z_2')$ space taken by the bins for equi-distant, equal-width bins in $z'$ with those (equal populated) bins defined in the main text.

\begin{figure}
    \centering
    \includegraphics[width=0.99\columnwidth]{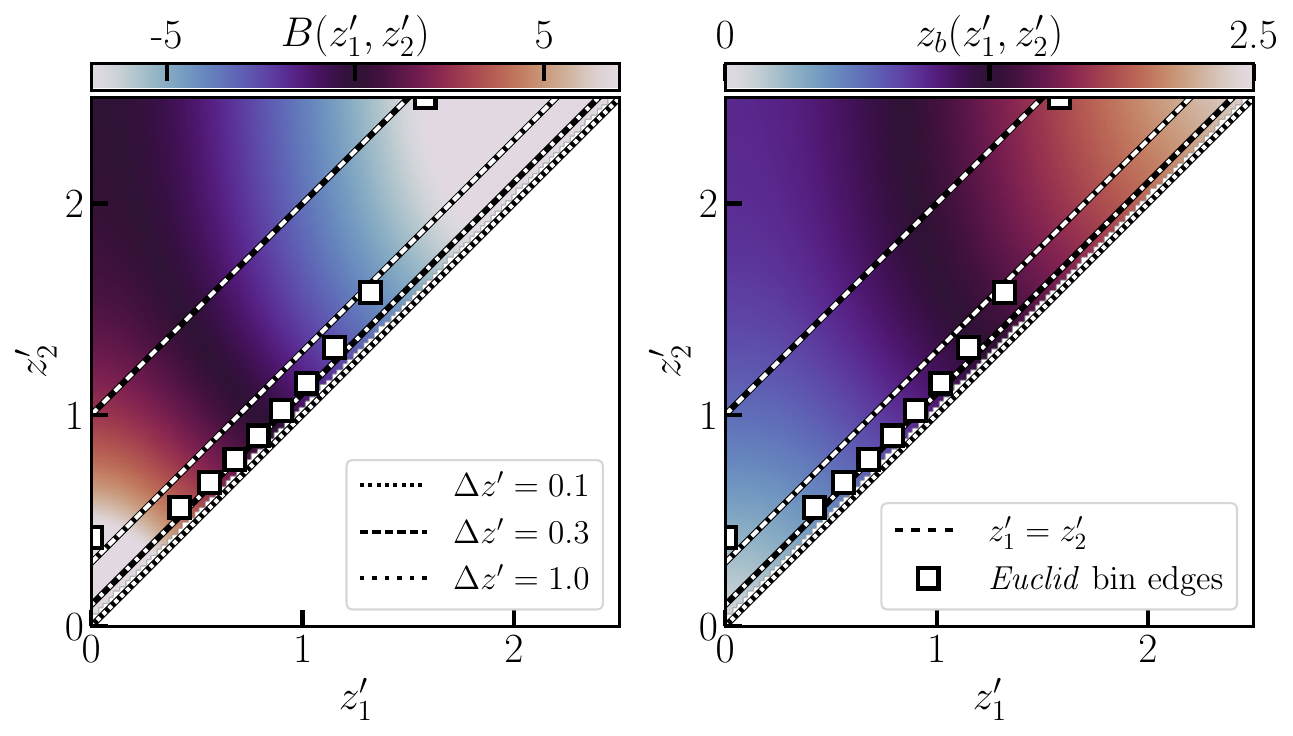}
    \caption{Boundary terms $B$ (\textit{left panel}) and effective redshift $z_b$ (\textit{right panel}) per choice of (photometric) bin edges $[z_1',z_2']$, computed according to  Eq.~\eqref{eq:B_Wb} and \eqref{eq:fz_zb_W}, and corresponding to the \textit{Euclid} specifications and Fig.~\ref{fig:Euclid_example}.  Dotted and dashed lines illustrate where equal-width bins would trace the functions $B$ and $z_b$.  Open squares show the deviation from the diagonals given by the equal-populated \textit{Euclid} bins defined in the main text.}
    \label{fig:zb_example}
\end{figure}

\begin{figure}
    \centering
    \includegraphics[width=0.6\columnwidth]{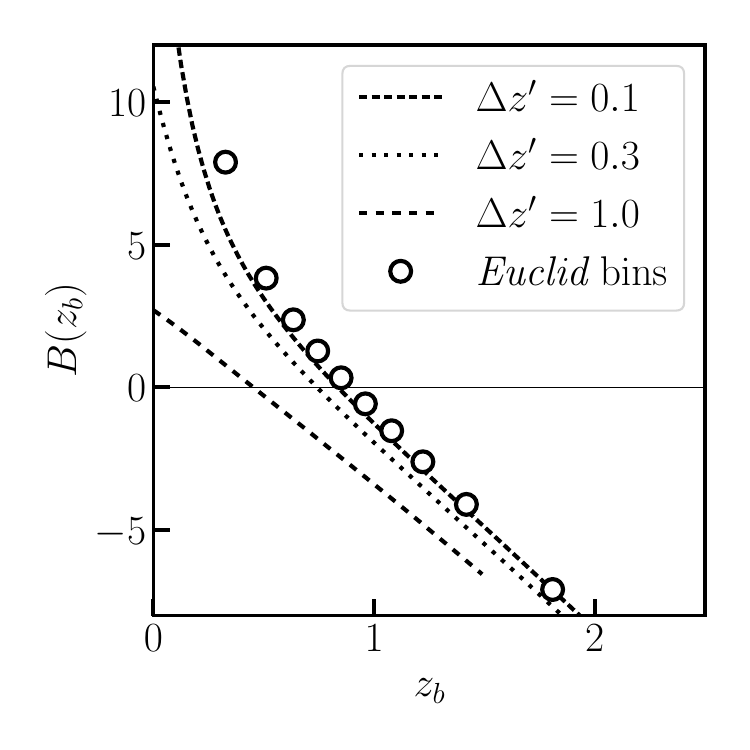}
    \caption{Boundary terms $B$ depending on effective redshift $z_b$ corresponding to (photometric) bin edges $[z_1',z_1'+\Delta z']$, as traced by the dotted and dashed lines in Fig.~\ref{fig:zb_example}.  Boundary terms are computed according to  Eq.~\eqref{eq:B_Wb} and \eqref{eq:fz_zb_W}.  With decreasing bin width different $B(z_b)$ asymptote to the same curve justifying the functions plotted in Fig.~\ref{fig:boundary1}.  A changing bin width deviates from this asymptote, albeit not by much. As examples, we show the values corresponding to the \textit{Euclid} specifications, Fig.~\ref{fig:Euclid_example}.}
    \label{fig:zb_example_2}
\end{figure}

Lastly, varying $\sigma_z$ for chosen narrow bins does not change the functional form of the boundary terms much, whereas the presence of redshift bias can. We discuss the influence of bias on the boundary terms in the next section.  However, we will only report the exact boundary terms at specific values of $z_b$, as opposed to drawing a full function, and will not continue the discussion on when $B$ as a function of $z_b$ is an appropriate approximation. Ultimately, boundary terms are only ever computed for a finite number of bins in any case.

\subsection{Bias studies}
\label{app:moreparameters_bias}

In addition to the simple checks in the main text, we here investigate the influence of bias and catastrophic outliers of redshift estimates for \textit{Euclid} on the boundary terms using a broader range of biases.\footnote{To keep the presentation contained we limit the discussion in this section to results for the \textit{Euclid} specifications.  However, the qualitative conclusions are equivalent also for \textit{Rubin}-LSST and SKA.} To this effect, consider the conditional distribution function~\citep[e.g.][]{2020A&A...642A.191E}
\begin{align}
    P(z',z) =& \phantom{+}\frac{1-f_o}{(2\pi\sigma_z^2)^{0.5}}\exp\left[-0.5\left(\frac{z'-mz-a}{\sigma_z}\right)^2\right]\nonumber\\
    &+ \frac{f_o}{(2\pi\sigma_z^2)^{0.5}}\exp\left[-0.5\left(\frac{z'-m_oz-a_o}{\sigma_{z,o}}\right)^2\right],
    \label{eq:P_bias}
\end{align}
where multiplicative bias $m$ was included in addition to the additive bias $a$ for the outlier (denoted by subscript $o$) and the non-outlier constituents of the sample. The redshift uncertainty may be different for the outlier subsample, which we denote by $\sigma_{z,o}$.  However, for the sake of studying arising trends, we consider \textit{all} redshifts to be biased in the same manner.  This stands in contrast to the main text, where we considered only $f_o=0.1$, i.e.~$10\%$ of the sample, the `outliers', are biased.  Here we set $f_o=0$ and consider multiplicative biases $m=\{0.9,1,1.1\}$, additive biases $a=\{-0.2,-0.1,0,0.1,0.2\}$, and changes to the uncertainty $\sigma_z=\{0.05,0.10\}$ (or equivalently, $f_o=1$ and the same sets of biases for $m_o$, $a_o$ and $\sigma_{z,o}$ while fixing $m=1$ and $a=0$).

As remarked already in the main text, as long as $P(z',z)$ has been inferred correctly from the data (including the contribution of the biases/outliers), Eq.~\eqref{eq:B_Wb} correctly computes the corresponding boundary terms.  However, if $P(z',z)$ is misestimated, e.g.~by not having included the outlier contribution, or the bias entirely, then Eq.~\eqref{eq:B_Wb} will not return the correct amplitude.  Furthermore, the estimation of $z_b$ that depends on $W_b(z)$ and with it on $P(z',z)$ itself will be influenced by biases (or their omission).  For the latter points, in the following, we present the respective bins' results centered on the true, \textit{unbiased} $z_b$ to highlight the differences in boundary terms that arise per bin.

\begin{figure}
    \centering
    \includegraphics[width=0.99\columnwidth]{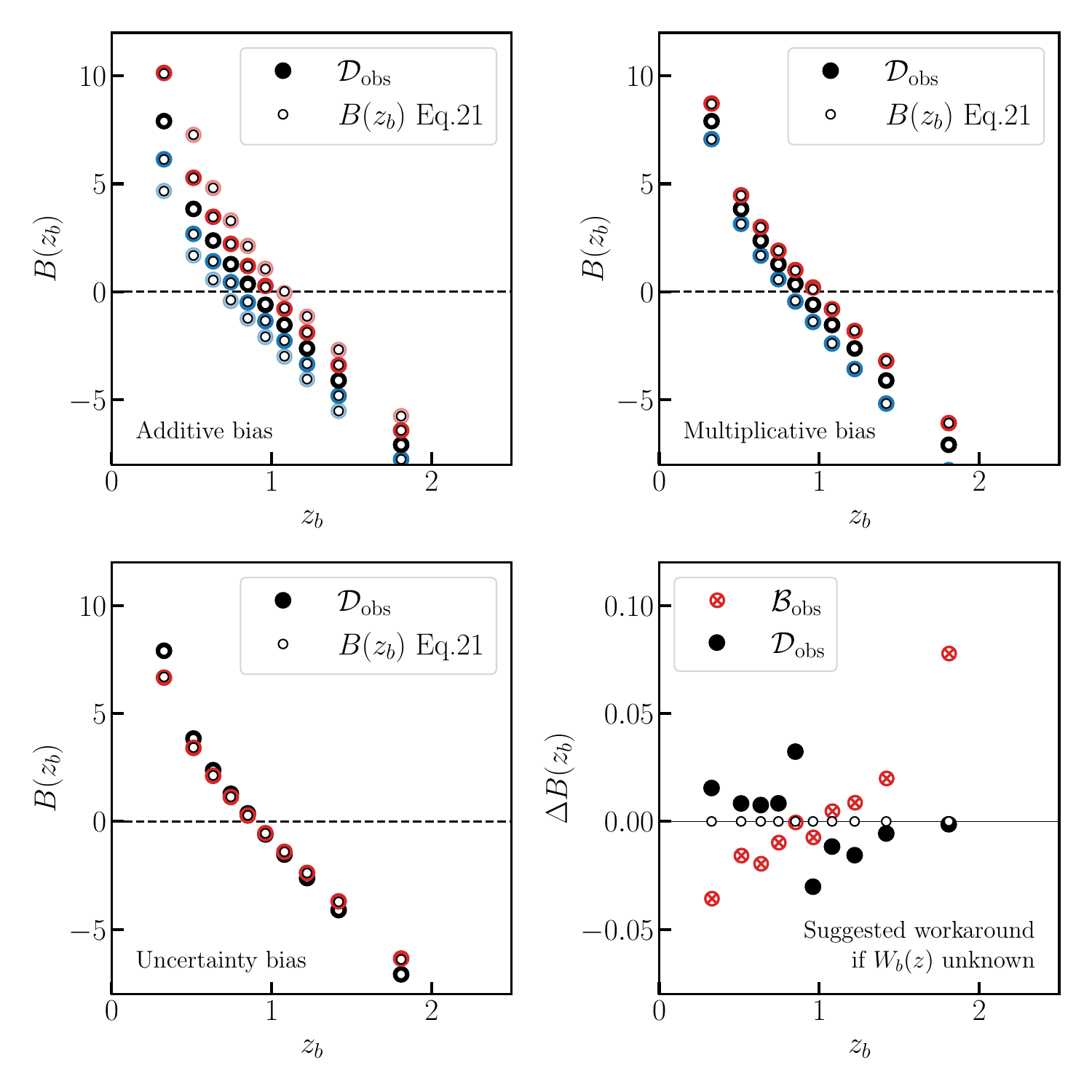}
    \caption{Boundary terms $B$ computed via Eq.~\eqref{eq:B_Wb} for \textit{Euclid} specifications and with consideration of various types of bias, as indicated by the labels in the bottom left (\textit{first three panels}).  These are compared against the median value of measured dipoles in 1000 mock samples, $\mathcal{D}_{\rm obs}$, demonstrating agreement between the theoretical and observed quantities.  If $W_b(z)$ is unknown, Eq.~\eqref{eq:B_tomography} offers an approximate solution by using the observed $f'(z')$ (\textit{last panel}).  Denoted~$\mathcal{B}_{\rm obs}$, $\Delta B(z_b)$ shows the differences between the observed quantities and the correctly predicted $B$ using Eq.~\eqref{eq:B_Wb} for the mildly-biased case shown in Fig.~\ref{fig:Euclid_biasexample}.}
    \label{fig:Euclid_biases}
\end{figure}

In the first three panels of Fig.~\ref{fig:Euclid_biases} we show the boundary terms for the same \textit{Euclid} specifications presented in the main text, where we consider each of the bias values separately.  There, we present as colored, filled dots the predictions of Eq.~\eqref{eq:B_Wb} using the correct values of the biases for $P(z',z)$.  To provide confidence in the accuracy of our analytical results, we also show, as empty dots, the median values of measured dipoles in 1000 random mock samples that were constructed based on the same biases.  We describe our simple simulations in section \ref{app:moreparameters_sampling} along with an efficient implementation of generating random samples from the \textit{Euclid} redshift distribution.  Demonstrated here, we see that Eq.~\eqref{eq:B_Wb} indeed correctly describes the boundary terms that would arise in a real sample.

Finally, we show in the last panel of Fig.~\ref{fig:Euclid_biases} the comparison of what we initially presented in Fig.~\ref{fig:Euclid_biasexample} with boundary terms computed directly from the distribution of (photometric) redshifts $z'$, $n'(z')$ and Eq.~\eqref{eq:B_tomography} using the known redshift boundaries.  This method does not explicitly compute $n(z)$ or $W_b(z)$ and therefore does not assume any biases, nor can it misestimate them.  For the present example, we find that $\mathcal{B}_{\rm obs}$ does not deviate much from the truth $B(z_b)$, when compared with the amplitudes expected from Eq.\,\eqref{eq:DEB}. However, in contrast to $\mathcal{D}_{\rm obs}$, $\mathcal{B}_{\rm obs}$ exhibits a trend in redshift when compared against $B(z_b)$.  This is unsurprising due to the bias that $z'$ carries with respect to $z$.  Since $n'(z')$ too is influenced by this bias, Eq.~\eqref{eq:B_tomography} will lead to results that is skewed with respect to the bias-corrected results of Eq.~\eqref{eq:B_Wb}. For larger biases than those considered here, the presented workaround solution will not be a viable option. Nevertheless, for a \textit{Euclid}-like experiment, we find that this method allows for the estimation of the rough magnitude of the boundary terms, even without explicitly modeling $W_b(z)$ or even knowing $P(z',z)$.  For the small biases expected in this context, this method may well prove useful.\\

In each of the cases we present only the median values of $\mathcal{D}_{\rm obs}$ and $\mathcal{B}_{\rm obs}$ were presented.  These values are influenced by the number of sources $N$ in each of the mocks, as well as by the total number of mocks and therefore generally have associated uncertainties. While the scatter seen in some of the panels is prone to these uncertainties, our purpose here simply is to highlight agreements or disagreements between the observed quantities and the theoretically predicted ones compared with their impact on the Ellis\&Baldwin formula Eq.~\eqref{eq:DEB}.  We therefore leave studying the impact of shot noise in number count maps and sampling errors of redshift distributions to future work.

\subsection{Mock data and Sampling aid}
\label{app:moreparameters_sampling}

We generate full-sky samples of $N$ data points isotropically distributed across the sphere.  Each point is assigned a cosmological redshift $\bar z$, sampled from Eq.~\eqref{eq:Smail}. Given each value of $\bar z$, (photometric) redshifts $\bar z'$ are sampled from the distribution~\eqref{eq:P_bias} with corresponding input parameters. Redshifts $\bar z'$ are then boosted via Eq.~\eqref{eq:z_boost} to give $z'$, and finally restricted by a selection function $W_b'(z')$. The finally obtained sample is then projected onto the sky to create a map of number counts.  The selection in boosted redshifts leads to a dipole in the map, that is measured via the least-squares \textsc{HEALPix}\footnote{\url{https://healpix.sourceforge.io}}~\citep{Gorski:2004by} routine \texttt{hp.fit\_dipole}, which is known to be unbiased for high source densities.  In our examples, we ensure this by selecting $N>10^6$ in each redshift bin.  The resulting mock maps are predicted to have dipole amplitudes exclusively described by $B(z_b)$ in the main text (as opposed to also respecting the terms seen in Eq.~\eqref{eq:DEB}), as neither aberration, nor Doppler boosting of flux densities is considered here.\\

Part of above steps is the generation of the redshift sample, which for the sizes $N$ that we attempt to reach requires an efficient sampling method. Since Eq.~\eqref{eq:Smail} is widely in use for modeling redshift distributions of upcoming surveys, we briefly describe our implementation.  Instead of rejection sampling the redshift distribution Eq.~\eqref{eq:Smail}, it is more efficient to perform inversion sampling.  We here present helpful expressions. The cumulative distribution function (CDF) $N(z)$ of Eq.~\eqref{eq:Smail} reads
\begin{align}
    N(z) = \int_0^z{\rm d}t\,n(t) = \int_0^z{\rm d}t\,t^2\exp\left[-(t/z_0)^\rho\right].
\end{align}
With the substitution $h\equiv(t/z_0)^\rho$, this can be written in terms of the lower incomplete gamma function
\begin{align}
    \gamma(s,x)\equiv&\int_0^x{\rm d}t\,t^{s-1}\exp\left[-t\right]\,,\label{eq:gammainc}\\
    N(z)=&\frac{z_0^3}{\rho}\gamma\left(\frac{3}{\rho},(z/z_0)^\rho\right)\,.
\end{align}
We then take the inverse of the CDF to find
\begin{align}
    N^{-1}(u) = z_0\left[\gamma^{-1}\left(\frac{3}{\rho},\frac{u}{z_0^3/\rho}\right)\right]^{1/\rho}\,,\label{eq:Ninv}
\end{align}
where $\gamma^{-1}$ stands for the inverse of Eq.~\eqref{eq:gammainc}.  A random sample of $n(z)$ is obtained via $N^{-1}(u)$ for a given sample $u$, where $u$ is randomly sampled from a uniform distribution $\mathcal{U}(0,1)$.  In practice, numerical implementations of inverse incomplete gamma functions, such as \texttt{scipy.special.gammaincinv()}, are comparably slow, so we first build an emulator/interpolator for Eq.~\eqref{eq:Ninv} before evaluating it for samples of $u$ -- once for every choice of parameters $(z_0,\rho)$.  Our emulator performs with a relative accuracy of $\lesssim10^{-8}$ and leads to $\sim150$ times faster sample generation compared with rejection sampling.

\section{Source selection according to color}
\label{app:otherselection}
In this appendix, we first detail why color cuts can be applied for sources obeying a perfect power law spectra without affecting the number count dipole. We then detail how the same cuts applied in conjunction with a running of the spectral index, generates additional contributions to the number count dipole.

\subsection{Color cut for power law spectra}
\label{app:otherselection_color}

It is useful to apply color cuts on sources to differentiate between objects of different nature such as stars or quasars, for example.  This was indeed done by \citeauthor{Secrest:2020has}, who imposed a color cut on WISE band magnitudes, $W_1 - W_2\geq 0.8$, to obtain a high-redshift sample of quasars with which they performed their measurement of the matter dipole.  In the present context, it might be suspected that the sharp color cut is accompanied by a smooth selection in redshift, see Sec.~\ref{sec:discussion}.  Here, we clarify that for the special case of power law spectra, we do not expect color cuts to generate additional contributions to the kinematic matter dipole.

Color, defined as the difference between apparent magnitude $m$ in two bands, A and B, reads
\begin{align}
C_{AB} \equiv m_A - m_B\,,
\end{align}
This may be immediately related to the absolute magnitudes $M$ and $k$-correction in each band
\begin{align}
C_{AB} = M_A - M_B + k_A(z) - k_B(z)\,,\label{eq:Color_M_k}
\end{align}
since the distance moduli to the same source cancel. We highlight here the dependence of the $k$-corrections on observed redshift $z$. This implies that in principle, a color cut corresponds to a certain cut on observed redshift, depending on how color and redshift are correlated, which, as we explain in Sec.~\ref{sec:discussion}, can generate what we call smooth boundary terms in Eq.\,\eqref{eq:Dkin_W}.  Following~\cite{Hogg:2002yh}, for a power law spectrum $S(\nu) \propto \nu^{-\alpha}$ and constant frequency/band-pass filters in the observer and source rest frames, we obtain the same $k$-correction for both bands $A$ and $B$
\begin{align}
k_{A/B}(z) = \frac{5(\alpha-1) }{2} \log_{10}(1+z)\,,\label{eq:D3}
\end{align}
where $z$ is the \textit{observed} redshift.  Because of this, the $k$-corrections in Eq.~\eqref{eq:Color_M_k} cancel, such that, for power law spectra, color is simply given by the difference in absolute magnitudes in each band $A$ and $B$. While these may be correlated with cosmic time (or cosmological redshift) for astrophysical reasons, we expect them to be independent of the observer's motion.  This implies that, in this case, a color cut, does \textit{not} generate an additional selection function, with which boundary terms could be associated.  Note that this does not hold in general, and nontrivial filters at the observer or a running of the power law spectral index are expected to generate a nontrivial dependence of color on observed redshift, leading to additional corrections to the Ellis and Baldwin formula.  We investigate the latter possibility in the following section.

\subsection{Running spectral index}\label{app:otherselection_running_spectral_index}

In the case where the power law spectrum has a running of the spectral index, the cancellation observed in the previous section need not hold.  Then, a color cut may affect the source counts differently in different directions relative to the direction of motion. This stems directly from the dependence of color on the observed redshift $z$ in \eqref{eq:Color_M_k} and \eqref{eq:D3}.  To show this, we follow the spirit of~\cite{Dalang:2021ruy,Lacasa:2024ybp} and calculate the contribution to the dipole by perturbing the rest frame source counts in terms of its variables.  Then, we apply the findings of~\cite{vonHausegger:2024jan} to show how evolution effects can be accounted for.

Suppose that the spectral index at the 2 frequencies corresponding to the 2 bands on which a color cut is imposed satisfy $\Delta\alpha = \alpha_A-\alpha_B \neq 0$.  This can happen if the spectrum is not very well modeled by a power law.  In this case, the number count as a function of redshift contains an extra term (as, e.g., compared to Eq.\,(25) of~\cite{Dalang:2021ruy}), which is the Taylor expansion on color cut $C_*$. While the color cut that the observer applies is independent of direction, it corresponds to different intrinsic color cuts depending on its redshift and direction as per the dependence of the $k$-correction on observed redshift in Eq.\,\eqref{eq:Color_M_k}.
\begin{align}
\frac{\p }{\p C_*} n_C(z, S_*,C_*) \delta C_*(z,\bs{\hat{n}})
\end{align}
where $\delta C_*(z,\bs{\hat{n}})= C_*- \bar{C}_* = 5 \Delta \alpha(z) \bs{\hat{n}} \cdot \bs{\beta} /(2\log(10))$ indicates the color cut difference with respect to the angular averaged color cut $\bar{C}_*$.  The only dependence on redshift can come from $\Delta \alpha(z)$, as can be computed from \eqref{eq:Color_M_k} and \eqref{eq:D3}.  One can define a $z$-dependent \textit{color bias}
\begin{align}
y(z)\equiv - \frac{5}{2}\frac{\p \log_{10} n_C(z, S_*,C_*)}{\p C_*}\,,
\end{align}
which captures the relative change in the cumulative number counts, as one varies the color cut $C_*$, as a function of observed redshift $z$.  In this situation, the matter dipole reads
\begin{align}
\mathcal{D}\e{kin}(z) & = \Big[ 3 + x(z) (1+\alpha_A(z)) \\
& \quad - y(z) \Delta \alpha(z) + \frac{{\rm d}\log n_C}{{\rm d}\log (1+z)}\Big] \beta\,.\label{eq:Dkin(r)_color}
\end{align}
where $\alpha_A$ appears because we assume that the flux cut is applied to sources in band $A$.  The kinematic dipole integrated over redshift then is computed as
\begin{align}
\mathcal{D}\e{kin} = \int_0^\infty \dd z\, f_C(z) \mathcal{D}\e{kin}(z)\,,\label{eq:Dkin_color_1}
\end{align}
where $f_C(z)$ is the normalized distribution of sources $n_C(z)$.  In~\cite{vonHausegger:2024jan}, it was shown that the first two terms reduce to the Ellis\&Baldwin formula
\begin{align}
\mathcal{D}\e{EB} = [2 + \t{x}(1+\t{\alpha}_A)]\beta 
\end{align}
with $\t{x}$ and $\t{\alpha}_A$ measured only on the sources which are close to the flux limit in the $A$ band.  In the present context, we specify that only sources with $C>C_*$ are considered for this (although both lower and upper color cuts could be included in the calculation naturally, as might be required for the analysis of Ref.~\cite{Panwar:2024xum}).  Following a similar spirit, one can measure $\Delta \t{\alpha}$ as the average on the sources which are close to the color cut, while simultaneously being above the flux cut $S>S_*$,
\begin{align}
\Delta \t{\alpha} = \int_0^{\infty} \dd z\, \t{f}_C(z) \Delta \alpha(z)
\end{align}
with 
\begin{align}
\t{f}_C(z) = \Big [\tilde n_C(z,S_*,C)\Big / \int_0^{\infty} \dd z\,  \tilde{n}_C(z,S_*,C)\Big]_{C=C_*}\,.
\end{align}
If one also defines an effective color bias 
\begin{align}
\t{y} & \equiv - \frac{5}{2}\frac{\p \log_{10} n_C(S_*,C_*)}{\p C_*}\,, \\
& = \int_0^{\infty} \dd z\, y(z) f_C(z)\,,
\end{align}
one can show that very similar cancellations happen as in~\cite{vonHausegger:2024jan}, such that Eq.~\eqref{eq:Dkin_color_1} reduces to 
\begin{align}
\mathcal{D}\e{kin} = \Big[ 2 + \t{x}(1+\t{\alpha}_A) - \t{y} \cdot \Delta \t{\alpha} \Big]\beta \,, \label{eq:Dkin_color_cut}
\end{align}
which is the kinematic dipole expectation for sources exhibiting power-law spectra with a running spectral index.  These quantities can in principle be measured with data:  While the color bias of, e.g., the sample of Ref.~\cite{Secrest:2022uvx} lies around $\t{y}\simeq 1$, the estimation of $\Delta\tilde\alpha$ or, generally, $\Delta\alpha(z)$ requires sufficient knowledge of the source spectra.  Spectral templates might enable a rough estimation; however, the advent of data from upcoming experiments will be required to empirically estimate the influence of the additional term derived here.

In summary, if the source spectra deviate from power laws, a color cut leads to an additional term in Eq.~\eqref{eq:Dkin_color_cut} with respect to the Ellis \& Baldwin formula, Eq.~\eqref{eq:DEB}.  Simultaneously, it must be respected that estimates of $\alpha$ using measurements in only two bands, $A$ and $B$, do not equate to the true $\alpha_A$ in band $A$, which is the relevant quantity if the flux cut is applied in that band, so also $\t{\alpha}_A\neq \t{\alpha}$ in general.  Depending on the particular source spectra these two effects might compete in increasing/decreasing the overall expected dipole amplitude.  We leave these detailed considerations for future work.  Lastly, the relation of this discussion to the formalism of this work, is made by the considerations in Sec.~\ref{sec:discussion}:  the additional contribution to Eq.~\eqref{eq:DEB} could have been computed also using smooth boundary terms.  The challenge there however is to find the corresponding selection function $W_b(z)$.

A relation between color cuts and magnification/evolution bias was explored recently in a different context by Ref.~\cite{2024arXiv241004705F}, who briefly hinted at the relevance this might carry for measurements of the matter dipole.  However, their discussions mainly take place in the context of spectra containing emission lines, where any such effect may well be more pronounced than in the case of smooth spectra.  Nevertheless, by having unveiled an explicit connection of redshift selections and boundary terms, our present work offers the framework for the precise computation of any corresponding contribution to the matter dipole amplitude.

\bibliographystyle{apsrev4-2}
\bibliography{References}

\end{document}